\documentclass{mn2e}
\usepackage{times}
\usepackage{graphicx}
\usepackage{lscape}
\usepackage{amsmath}
\usepackage{amssymb}
\usepackage{epsfig}
\usepackage{psfrag}
\usepackage{longtable}
\usepackage{dcolumn}
\newcolumntype{d}[1]{D{.}{.}{#1}}

\newcommand{\etal}{{et al}\/.}

\title[Bayesian inference of jet bulk-flow speeds in FRII radio sources]{Bayesian inference of jet bulk-flow speeds in FRII radio sources}
\author[L. M. Mullin \& M. J. Hardcastle]{L. M.
  Mullin$^{1}$\thanks{E-mail:
lmm37@hotmail.co.uk} and M. J.
  Hardcastle$^{2}$\thanks{E-mail: m.j.hardcastle@herts.ac.uk. Address
    postal correspondence to MJH.}\\
$^{1}$Astrophysics Group, Cavendish Laboratory, University of Cambridge, J J Thomson Avenue, Cambridge CB3 0HE\\
$^{2}$School of Physics, Astronomy and Mathematics, University of Hertfordshire, College Lane, Hatfield AL10 9AB}
\begin{document}

\date{Draft of \today}



\maketitle

\label{firstpage}

\begin{abstract}
Radio jet and core data for a complete sample of 98 FRII sources with
$z < 1$ are analysed with a Markov-Chain Monte Carlo (MCMC) model
fitting method to obtain constraints on bulk-flow speeds in the beam.
The Bayesian parameter-inference method is described and demonstrated
to be capable of providing meaningful constraints on the Lorentz
factor at both kiloparsec and parsec scales. For both jets and cores
we show that models in which some intrinsic dispersion is present in
the features' intrinsic prominence, bulk-flow speeds or both provide
the best fit to the data. The constraints on the Lorentz factor on
parsec scales are found to be consistent with the expected values
given VLBI observations and other evidence, with $\bar\gamma
  \approx 10$--$14$. On kiloparsec scales, the Lorentz factor is
found to be $\approx 1.18$ --$1.49$, in agreement with the
results of previous analyses of radio jet data. These values are
clearly not consistent with the $\gamma \approx 10$ speeds required by
beamed inverse-Compton models of X-ray emission from quasar jets; our
results therefore support models that require velocity structure in
powerful jets.
\end{abstract}


\section{Introduction}
\subsection{Relativistic beaming of jet emission}
\label{sec:history} 
Fanaroff \& Riley (1974) type II quasars and radio galaxies (hereafter
FRIIs) generally exhibit double radio lobes with scales of tens to
hundreds of kiloparsecs that are symmetrical about the source's
central engine. There is now a very large amount of evidence for
models (e.g. Scheuer 1974; Blandford \& Rees 1974) in which these are
supplied with energy, mass, momentum and magnetic flux by a bipolar,
symmetrical, continuous flow of material -- the `beam' or `beams'. These outflows
must persist and be well-collimated out to the 100-kpc scales of the
lobes in order to give rise to the observed compact terminal hotspots.
However, the observational signatures of these collimated outflows,
`jets' (e.g. Bridle \& Perley 1984), are not always detected in FRIIs,
either in the radio or at other wavebands. Where jet emission is
observed, the jet is very often `one-sided': i.e., it is either
detected on one side of the source only or is very much brighter on
one side. This is particularly characteristic of powerful FRII
quasars. As the presence of twin beams is suggested by the morphology
of the large scale structure, the fact that the detection of both jets
in these powerful sources is so uncommon supports the hypothesis that
the beam's emitting material is moving at relativistic speeds on kpc
scales and that the emission is affected by Doppler boosting
(`beamed'). For the case of a source where the beam axis makes a
relatively small angle to the line of sight, as would be the case in
standard unified models (Barthel 1989) for an FRII quasar, emission
associated with the approaching jet is then Doppler-boosted and
observable, while that associated with the receding jet is
Doppler-suppressed and not detected. For a source where the beam axis
lies close to the plane of the sky, both jets are likely to be
Doppler-suppressed.

The key piece of evidence for relativistic beaming on kpc scales comes
from the tendency for the jet side to be associated with the less
depolarised lobe of the source, the Laing-Garrington effect
(Garrington et al.\ 1988, Laing 1988). The depolarizing mechanism is
believed to be an external Faraday screen in which the source is
embedded, presumably the hot phase of the intergalactic medium, and
thus the degree of depolarization observed depends upon the path
length of the radiation through the screen. The less depolarized lobe
is therefore expected to be the nearer lobe, or the lobe pointing
towards us, and any correlation with the kiloparsec jet requires that
the jet emission is Doppler-boosted. However, while observations of the
Laing-Garrington effect require relativistic jet speeds on kiloparsec
scales, they do not tell us what these speeds are.

Direct evidence for relativistic flow speeds in the {\it inner}
regions of the beams, on parsec and sub-parsec scales, comes from VLBI
observations of apparent superluminal motion. For example, Hough
\etal\ (2002) have mapped the parsec scale regions of a complete
sample of 25 lobe-dominated quasars (defined as having a ratio of
nuclear to extended flux density at 5~GHz of less than 1) from the
sample of Laing, Riley \& Longair (1983, hereafter LRL). Of these
sources, all have resolved structure on parsec scales, with one-sided
jets detected on the same side as that of the kiloparsec jet where
such a jet is detected. Hough \etal\ estimate bulk flow speeds with
Lorentz factor, $\gamma$, $\approx 5-10$ from multi-epoch
observations, and these results are consistent with those from other
observing programmes. Further observational support for the idea of
high bulk speeds in radio-loud AGN in general comes from the rapid
variability and consequent high brightness temperatures of
parsec-scale features, the absence of very strong inverse-Compton
emission in X-ray observations of the nucleus, and transparency to
high-energy $\gamma$ rays, which together imply bulk Lorentz factors
$\gamma\ge2$ and possibly as high as $\sim 50$ (Begelman, Fabian \&
Rees 2008). Arguments based on unification and population statistics
(mostly in low-power objects; e.g. Chiaberge \etal\ 2000, Hardcastle
\etal\ 2003) imply $\gamma \sim 3$: a plausible explanation for the
widely differing $\gamma$ values is that there is velocity structure
in the parsec-scale jet. However, the essential point for our purposes
is that all parsec-scale estimates agree on the need for Lorentz
factors corresponding to speeds greater than $\sim 0.9c$, and the
direct VLBI estimates imply speeds $\ga 0.99c$ in general. It is
therefore important to ask whether these speeds persist to the
kiloparsec scale.

Two approaches to determining the kiloparsec-scale jet speed have been
taken in the literature. The first uses the radio properties of the
jets. This approach was pioneered by Wardle \& Aaron (1997, hereafter WA97), who
analysed the observed jet flux asymmetry in the 13 quasars imaged by
Bridle \etal\ (1994). The jet flux asymmetry was defined as the
observed jet flux over the counter-jet flux (where the counter-jet is
the fainter of the two), but as for most sources no counter-jet was
actually detected, many of their data points are actually lower
limits. Taking into account the possibility of some intrinsic
asymmetry, they simulated a number of data sets that were compared to
the observed data by means of a Kolmogorov-Smirnov (K-S) test. They
found that the observed data are best fitted with $0.6 \le \beta \le
0.8$, where $\beta$ is the jet speed as a fraction of the speed of light. However, because their parent sample was not complete, they were
forced to take quite a complex approach to the inclusion of the
selection criteria in their analysis, and effectively to treat the
upper limit on angle to the line of sight made by the beam axes of
their sources as a free parameter in their fits. Hardcastle et al.
(1999, hereafter H99), used a similar method to constrain the jet
bulk-flow velocities for their sample of FRIIs with $z \le 0.3$, which
overlaps considerably with the sample considered in the present paper
(see Section \ref{sec:sample}). Rather than using jet sidednesses they
used the jet and core prominences (defined as the ratio between the
jet/core flux density and that of the extended emission: see Section
\ref{sec:promdef}). Because their sample was complete the source
orientation could be assumed to be random, simplifying the analysis
with respect to WA97's work. In their analysis the free
parameters were the intrinsic prominence, $p_{\rm int}$, and $\beta$ and, exploring a grid in
these two parameters and using K-S tests in the same way, they derived
speeds between $0.5c \sim 0.7c$. Arshakian \& Longair (2004) used an
analytic approach to the H99 data to infer slightly lower speeds,
$\beta \approx 0.4$, on the basis of the jet sidedness distribution,
while constraining $\beta > 0.6$ for the sample used by WA97. Thus all the approaches based on the distribution of the
observed properties of the radio jets to date have been consistent in
implying only moderately relativistic bulk speeds, $\beta \approx 0.5
\pm 0.1$.

However, a different approach is motivated by the widespread detection
of strong X-ray emission from the jets of core-dominated quasars,
believed to be the highly aligned counterparts of the FRII radio
galaxies and lobe-dominated quasars studied by WA97
and H99. Following the discovery of the prototype of this class, PKS
0637$-$752 (Schwartz \etal\ 2000), it was quickly realised that the
broad-band spectra of these object preclude a one-zone synchrotron
model for the radio through X-ray data, while various inverse-Compton
models for the X-rays require extreme departures from equipartition
for a non- or mildly relativistic jet. Instead, the model proposed
independently by Tavecchio \etal\ (2000) and Celotti, Ghisellini, \& Chiaberge (2001) is
widely adopted. In this model, the jet is moving relativistically,
with a bulk Lorentz factor $\gamma \gg 1$. As seen by the jet, the
energy density in the microwave background increases by a factor of
the order $\gamma^2$, and this increases the emissivity of the
inverse-Compton scattering of the microwave background (hereafter
CMB/IC) in the jet frame; in PKS 0637$-$752, crucially, the $\gamma$
($\sim 10$) required for the kpc-scale jet is very similar to that
inferred from VLBI studies of superluminal motion in the nucleus. The
emission from this process is strongly anisotropic and so is only
visible in core-dominated objects, but, in unified models, the jets in
lobe-dominated quasars and FRII radio galaxies must have comparable
speeds. The implication of $\gamma \gg 1$ in the kpc-scale jets of all
radio-loud objects is in strong contrast to the results of the
prominence/sidedness analyses described above.

How can these two very different estimates of the kpc-scale jet speed
be reconciled? There is still disagreement in the literature over
whether the beamed CMB/IC model really does describe all, or even any,
of the observed quasar jets (e.g. Stawarz \etal\ 2004, Hardcastle
2006, Jester \etal\ 2007). If it does, then, as argued by Hardcastle
(2006), velocity structure in the kiloparsec-scale jets and perhaps
bulk deceleration on hundred-kpc scales seem inevitable consequences.
At the same time, though, the existing work on the radio data is open
to a number of criticisms. The K-S test is not really adapted to model
fitting (that is, it is not obvious that maximizing the K-S test null
hypothesis probability really corresponds to maximizing the
likelihood). More seriously, the large number of jet or counterjet
non-detections and consequent limits on sidedness or prominence
measurements are hard to take into account either in the K-S method of
WA97 or H99 or in the analytical method of Arshakian \&
Longair (2004), although H99 attempted to assess the effect of the
limits in their sample by scaling them and argued that they did not
have a strong effect. In all cases the sample sizes are small;
additionally, WA97's sample has complex selection effects
while H99's sample is low-luminosity and contaminated by
low-excitation radio galaxies whose role in unified models is not
clear. For all these reasons, it is worth revisiting the radio-based
estimates of kpc-scale jet speeds with new data and new analysis
techniques: the present paper presents the results of such a study.

\subsection{This paper}

The present paper is based on the work of Mullin, Riley \& Hardcastle
(2008, hereafter Paper I). In that paper we presented a detailed study
of the observed properties of a complete sample of FRII sources,
including the kiloparsec-scale jets and core features. We concluded that
the observational evidence supports the beaming hypothesis and that,
while there is stronger evidence for relativistic speeds on parsec
scales, the observed correlation between jet and core brightness
implies the extension of high bulk-flow speeds into kiloparsec scales.

In this paper we use these jet and core data together with a Bayesian
inference method in order to constrain Lorentz factors in the beams on
parsec and kpc scales. Our approach is free from many of the
disadvantages of earlier work. We do not carry out systematic grid
searches of parameter space, and so are not limited to a small number
of model parameters: this allows us to deal with the case in which the
intrinsic prominence and speed distributions are not delta functions
but themselves have some intrinsic scatter. Crucially, we can also
treat the limits in the data properly, rather than treating them as
measurements. In addition, our dataset is a factor $\sim 2$ larger
than that of H99, and contains luminous quasars and powerful radio
galaxies which are well matched to the core-dominated quasars for
which high bulk Lorentz factors have been inferred. This paper
therefore represents a significant improvement over previous work.

The remainder of the paper is structured as follows. In the following
Section the dataset is described, while the analysis method and
performance of the code used is discussed in Section \ref{sec:models}.
Results are presented in Section \ref{sec:results} and the discussion
and conclusions are in Section \ref{sec:conclusions}.

Throughout the paper we use the quantities measured and calculated in
Paper I: this implies the use of a cosmology with $H_0 = 70$ km
s$^{-1}$ Mpc$^{-1}$, $\Omega_{\rm m} = 0.3$ and $\Omega_\Lambda =
0.7$. All symbols used are summarized in Table \ref{tab:gloss}.

%
%
\section{The data}
%
\subsection{The sample}\label{sec:sample}
The sample is that of Paper I, which consists of the 98 FRII radio galaxies and quasars with $z<1$ in the sample of LRL. This is a complete flux-limited sample, including all 3CR sources observed to have $S_{\rm{178}} > 10.9$ Jy (on the scale
of Baars et al., 1977) with declination $> 10^\circ$ and $ |b| >
10^\circ$, where $S_{\rm{178}}$ is
the total source flux measured at 178~MHz. At this low frequency the
source flux is dominated by the diffuse emission of the large scale
lobe structure and as such little contribution should be made by
Doppler-boosted components; thus the selection criterion should ensure
that the sample is not biased with respect to orientation.

A long-term observing project has mapped the vast majority of these 98
sources at high resolution and sensitivity with the VLA telescope.
Data from this project have been presented in a series of papers:
Black et al. (1992), Leahy et al. (1997), Hardcastle et al. (1997),
Gilbert et al. (2004) and, most recently, Mullin, Hardcastle \& Riley
(2006). Observations of the outstanding sources have been made by us
or by other workers and the data are available in the literature; all
references to the data used here are given in Paper I. In addition,
the data -- maps and measurements -- from Paper I are now available
on-line\footnote{See http://zl1.extragalactic.info/}. The subsample of
sources with $z\le 0.3$ overlaps considerably with that of H99, and
the measurements of H99 are used for the sources that we have in
common, but the larger redshift range of our sample gives us a factor
2 more sources (improving the statistical significance of our results)
and means that 15 FRII quasars are included, complementing the data on
broad- and narrow-line radio galaxies from H99's work.
%
\subsection{Jets}\label{sec:jetdef}
The definition of the term jet and a discussion of the data and measurements is given in Paper I. Here, the jet criteria are reiterated: they are based on those of Bridle \& Perley (1984). Thus, a jet is any feature that is
\begin{enumerate}
\item at least four times as long as it is wide;
\item separable at high resolution from other extended structures (if
any), either by brightness contrast or spatially (e.g. it should be a
	      narrow ridge running through more diffuse emission, or a
	      narrow feature in the inner part of the source entering
	      more extended emission in the outer part).
\end{enumerate}
In some sources jets appear to bend, and this causes problems for
analysis in terms of beaming models, which must assume a single angle
to the line of sight $\theta$, as discussed by Bridle et al. (1994).
We follow Bridle \etal\ (1994) and H99 in defining the straight jet,
which satisfies the above two criteria but also must be aligned with
the compact radio core where it is closest to it (and is measured from
the end closest the core along its length only while the deviation
from a straight line is less than the jet radius). Only the flux
density of the straight jet is used in the analysis in this paper.

In practice, the straight jet is taken to be the longest straight
section of the jet in the source that is aligned with the core. Using
the {\sc aips} task TVSTAT, the integrated flux within the region
containing the apparent jet emission, $F_{\rm obs}$, was found.
Background flux was corrected for by taking measurements of two
regions identical in size to the initial jet measurement to the
immediate right and left of the feature. The average of these, $B_{\rm
obs}$, was then subtracted from the jet measurement to give the
observed jet flux, $J_{\rm obs}=F_{\rm obs}-B_{\rm obs}$. In order to
get the best estimate of $J_{\rm obs}$, three values of jet flux were
taken this way and averaged. The greatest source of error in the jet
measurement is considered to arise from the ambiguity in defining the
jet emission itself: the errors quoted are therefore based on the
range of the three jet measurements made. Where no jet emission is
detected, an upper limit is estimated by measuring the integrated flux
of a region $\approx2$ restoring beam widths across the entire
distance between the core and primary hotspot region. Background flux
is corrected for in the same manner as for the detected
jets by taking two further integrated flux measurements either side of
the initial region. However, if the flux associated with the central
region is not the highest of the three, then the upper limit estimate
is taken to be the positive difference between the central measure and the lower
of the other two.

The sample extends over a large range in redshift and was
  observed using a variety of different telescopes and telescope
  configurations, which means that the effective (spatial) observing
  resolution is far from constant across the sample. Observational
  effects on jet detectability were considered in Paper I, where we
  concluded that, although observing resolution is clearly a factor in
  jet visibility, there is no simple {\it systematic} bias across the
  sample nor any trend with redshift. We therefore do not expect that
  the variations in effective observing resolution will affect the
  robustness of the results of any analysis of $p_{\rm obs_{j}}$.
%
\subsection{Cores}\label{sec:coredef}
Paper I also contains a detailed discussion of core measurements. The core measurements were obtained from the highest-resolution map available for a given source using the {\sc aips} task JMFIT, which fits an elliptical Gaussian model of between one and four components to a feature. One component was fitted and the peak intensity found was taken as the core flux. As most cores in the sample were unresolved at all resolutions such a model fitted the data well. Errors were determined from the square root of the average of the squared formal errors returned from the fitting procedure. For around two thirds of the sample this error is less than 2 per cent of the core flux, so the calibration error (expected to be 2-3 per cent) will dominate. Errors quoted therefore correspond to 3 per cent of the core flux measurement, unless the formal error from JMFIT is greater, in
which case the latter is quoted. Cores were undetected in only seven
sources, and in these cases estimates of upper limits were made from
the off-source noise. 

Again, observing effects were considered for the core prominence
  data in Paper I. We concluded that, as the cores are typically
  bright, unresolved features that are generally much brighter than
  any lobe or jet material with which they might be confused at low resolution, observational effects are not a source of systematic bias in the analysis of $p_{\rm obs_{c}}$. 

%
\subsection{Prominence}\label{sec:promdef}
In order to analyse the jet and core data the observed jet and core flux density data are extrapolated to a common frequency of 8.4 GHz and then K-corrected assuming the spectral index, $\alpha, =0.5$ for jets and $=0$ for cores, before being converted to a luminosity using the relation
%
%
\begin{equation}
P = R^{2} (1 + z)^{2} S
\end{equation}
%
%
where $P$ is the jet or core luminosity, $R$ the proper distance (as calculated using the {\sc angsiz}\footnote{http://ascl.net/angsiz.html} code) and $S$ the jet or core flux density. The total source flux measured at 178 MHz is K-corrected using the low-frequency spectral index, $\alpha_{lf}$ (appropriate over the range 178 -- 750~MHz), and converted to a luminosity, $P_{178}$. The jet and core prominence are then the ratio of the corresponding feature's luminosity to $P_{178}$. Observed jet and core prominences for our sample
are plotted in Figs \ref{fig:jetmodel} and \ref{fig:coremodel}.

This definition of prominence, which we used in Paper I, assumes
  that the total source flux as measured at 178 MHz is uncontaminated
  by beamed components. It is expected that the sample will be
  dominated by the extended lobe emission at this low frequency;
  however, in sources with very bright jet or core features it is not
  clear that no contamination exists and this is potentially a source
  of bias in the prominence values. We investigated whether there is
  any evidence for such contamination in our data by comparing
  prominence values evaluated as described above with those evaluated
  with a modified total source flux, that is, a total source flux
  corrected by subtracting off jet and core features.

The modified total source flux at 178 MHz was calculated by extrapolating the jet and core fluxes from the observed frequency to 178 MHz. Here, the {\it total} jet was used, that is, the feature that satisfies the jet conditions as specified in Section \ref{sec:jetdef} and not the more restrictive {\it straight} jet conditions (see Paper I for a more detailed description of total and straight jet definitions). Both jet and counter-jet features were extrapolated back to 178 MHz. The corrected fluxes were then subtracted from the total source flux, and this modified total source flux was K-corrected and used to evaluate the core and straight jet prominence as described above.

Plotting these alternative jet and core prominence values against those initially determined, there was no evidence that either 
the core or jet prominence values were greatly affected by
contamination of beamed components if the total source flux at 178 MHz
is used without first correcting it -- the correlation in the plots is linear and there is no trend for sources to curve away from this line. Additionally, considering the errors in the jet and core flux measurements, the difference in prominence values made by modification of the total source flux is low -- in particular, the difference in the jet prominence made by modifying the total source flux is typically a fraction of a percent of the quoted error. 

From this we conclude that using the 178 MHz total source flux
without making any attempt to remove beamed emission to
define prominence, as we did in paper I, will not affect the
robustness of our analysis. We therefore use the
unmodified prominence for consistency with Paper I.

Finally, we note that in Paper I a spectral index of 0.5 was used for jets,
despite the fact that higher values of $\approx 0.8$ have typically
been used by other workers (H99, for example). Hotspot features are
expected to be associated with much flatter spectra than jets and it
was decided in the previous analysis, given that trends in core, jet
and hotspot properties as well as correlations between the features
were being considered, that a spectral index of 0.5 should
be used for both jet and hotspot features. The analysis presented here
uses the jet prominence data of Paper 1; however, we consider the
effects of varying $\alpha$ for jets in Section \ref{sec:jets}.
%
\section{Data models}\label{sec:models}
%
\subsection{Doppler boosting of source emission}\label{sec:beamingmod}
The emission from any component of a radio source that is travelling
at a significant fraction of the speed of light with respect to an
Earth-bound observer will be anisotropic due to relativistic beaming
even if it is isotropic in the rest frame (in this paper we neglect
the minor effects due to intrinsic anisotropy of emission in jets: see
H99 for more discussion). As received on Earth, a feature will have an
observed flux density, $S_{\nu_{\rm obs}}$, given by
%
%
\begin{equation}
S_{\nu_{\rm obs}} = S_{\nu_{\rm rest}}(\gamma [1 - \beta \cos \theta])^{-m+\alpha}
\label{eq:beaming}
\end{equation}
%
%
(e.g. Ryle \& Longair 1967; Bridle et al. 1994) where $S_{\nu_{\rm rest}}$ is the flux density of the feature in the emitter's rest frame, $\beta$ is the fraction of the speed of light at which the emitter is traveling, $\gamma$ is the Lorentz factor ($=1/\sqrt{(1-\beta^{2})}$), $\theta$ is the angle of the
velocity vector to the line-of-sight, $\alpha$ is the spectral index
of the radiation (where $S \propto \nu^{-\alpha}$) and $m$ is a
constant reflecting the geometry of the beamed component. Following
Scheuer \& Readhead (1979) and Lind \& Blandford (1985), the value of
$m$ taken here to be appropriate for a continuous jet is 2.  As
  the spectral index appears in the exponent of the Doppler factor, 
  it follows that, all other things being equal, a larger
  assumed $\alpha$ gives rise to a stronger beaming effect.
\subsection{Jets}\label{sec:jetmodels}
Applying equation (\ref{eq:beaming}) to the jet features in the
sample, we obtain the relationship between the observed jet prominence, $p_{\rm obs_{j}}$, and the intrinsic jet prominence, $p_{\rm int_{j}}$:
%
%
\begin{equation}
p_{\rm obs_{j}} = p_{\rm int_{j}}[\gamma(1 - \beta \cos \theta)]^{-2 +
\alpha} \label{eq:jet_model}
\end{equation}
%
%
where $\alpha$ is the adopted spectral index of the jet, $p_{\rm int_{j}}$ is the intrinsic, rest-frame prominence of the jet
feature and $\theta$ is the orientation of the jet with respect to the
observer's line of sight ($\gamma,\beta$ are as before). Here we
assume that the normalizing luminosity ($P_{178}$ in our case) is
unaffected by beaming, as discussed above.
%
\subsection{Cores}\label{sec:coremodels}
The core emission is thought to originate in the inner parsecs of the beam, that is, the parsec-scale bipolar jets, and so the model of equation (\ref{eq:beaming}) can be applied to the observed core prominence in the following manner, assuming that for cores $\alpha=0$: 
%
%
\begin{eqnarray}
p_{\rm obs_{c}} & = & p_{\rm int_{a}}[\gamma_{\rm a}(1 - \beta_{\rm a} \cos \theta_{\rm a})]^{-2}\notag\\
            &&  + p_{\rm int_{r}}[\gamma_{\rm r}(1 - \beta_{\rm r} \cos \theta_{\rm r})]^{-2} \label{eq:core_model}  
\end{eqnarray}
%
%
where the subscripts `a' and `r' correspond to the approaching and
receding parsec scale jet respectively. Again the subscript `int' indicates the
intrinsic, rest frame prominence of these jets and $\theta$, $\gamma$
and $\beta$ are as before. In the simplest models, which we 
adopt throughout the paper, $p_{\rm int_{a}} =
  p_{\rm int_{r}} = 0.5 \times p_{\rm int_c}$, $\theta_{\rm r} =
180^\circ - \theta_{\rm a} = 180-\theta$ where $\theta$ has the same
value as for the kpc-scale jet, $\beta_{\rm r} = \beta_{\rm a} =
\beta_{\rm c}$ and
$\gamma_{\rm r} = \gamma_{\rm a} = \gamma_{\rm c}$.
%
\subsection{Model Fitting Method}\label{sec:analysis_method}
%
\subsubsection{Beaming Model Validity}\label{sec:modelval}
As discussed in Section \ref{sec:history}, previous workers have used
jet and core models of the form given in equations
(\ref{eq:jet_model}) and (\ref{eq:core_model}) to constrain $\gamma$
values. An advantage of the present analysis method over these studies
is the possibility to allow for distributions in the intrinsic
prominence and $\gamma$ parameters. Arguments have been made (e.g.
Urry \& Shafer 1984) that the intrinsic jet or core prominence might
be expected to be a fixed fraction of the intrinsic total source flux,
but, while this is a useful simplifying assumption, it is more likely
that there will be a range in intrinsic prominence values in the
sample, since if nothing else there will be scatter in the
relationship between the total (normalizing) luminosity and the
intrinsic jet power. 

In the present paper the nature of such distributions is assumed
  to be normal or log-normal in the case of jets, since such
  distributions are appropriate to cases in which the observed values
  are the sum, or product, of many variable factors. This is also the
  case for the cores, but in this case we also consider a model based
  on power-law distributions. There have been a number of studies in
  the literature of samples of VLBI observations in terms of the
  apparent velocities and luminosities, often including model fitting
  to obtain information on the probable $\gamma$s and luminosity
  functions. In these studies, a power-law distribution for $\gamma$
  is often assumed. For example, Urry \& Padovani (1990) consider
  the effects of allowing such a distribution in the bulk Lorentz
  factors in their analysis, arguing that it allows both low or high
  values of $\gamma$ to be favoured, but also that a wide Gaussian
  distribution would resemble a flat power-law. Subsequent
  studies, such as Lister \& Marscher (1997), and more recently Cohen
  \etal\ (2007), find that model fitting based on such distributions is
  consistent with the data, though no evidence that such a model is to
  be {\it favoured} is reported. Given this other work, applying a
  power-law distribution to the core data provides an interesting
  comparison to models based on a normal distribution. A uniform
  distribution in $\cos \theta$ is used for $\theta$, assuming that
  the sample sources are randomly aligned with respect to the
  observer's line of sight.

\begin{figure}
\epsfxsize 8.5cm
\epsfbox{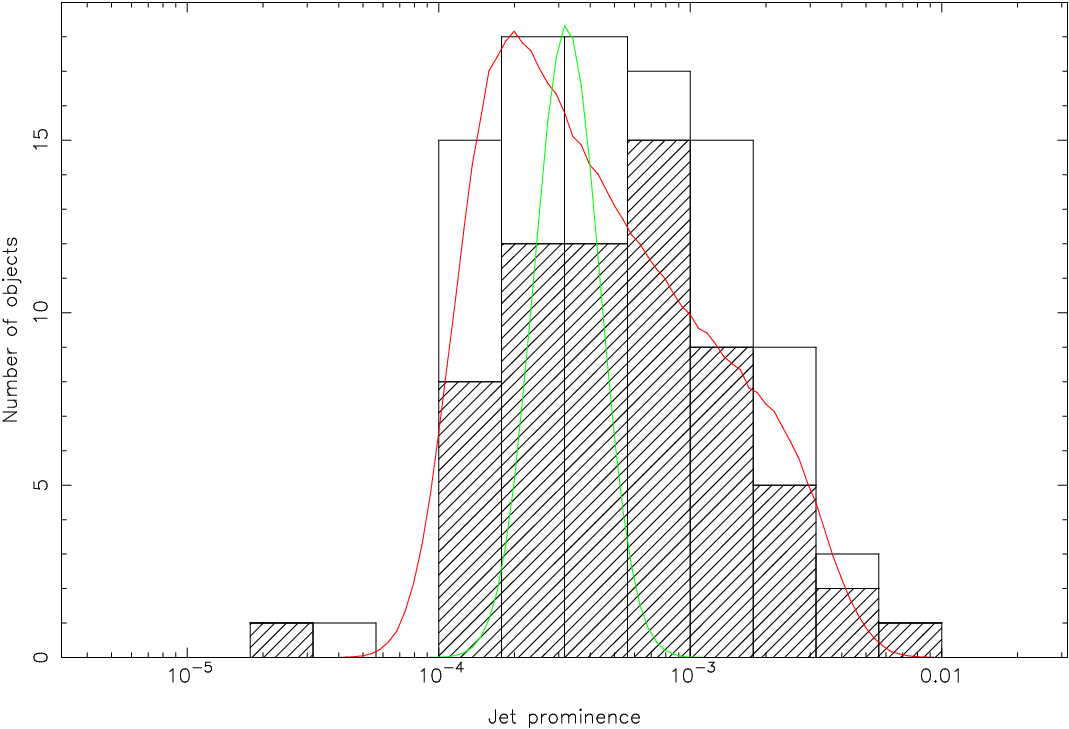}
\caption{\label{fig:jetmodel}Histogram of the observed jet prominence
  data, $p_{\rm obs_{j}}$. Filled regions indicate measurements and
  empty regions upper limits. The data are compared (solid lines) with the expected
  distribution jet prominence from a simulation with $\gamma = 1.5$
  (broad curve, red) and the corresponding distribution of intrinsic
  prominence (narrower curve, green), renormalized to a common
  maximum value for convenience of presentation.
  $p_{\rm int_{j}}$ has a lognormal distribution with $\sigma=0.3$ (in
  units of the natural logarithm) around $\ln(p_{\rm int_{j}}) = -8$.}
\end{figure}

\begin{figure}
\epsfxsize 8.5cm
\epsfbox{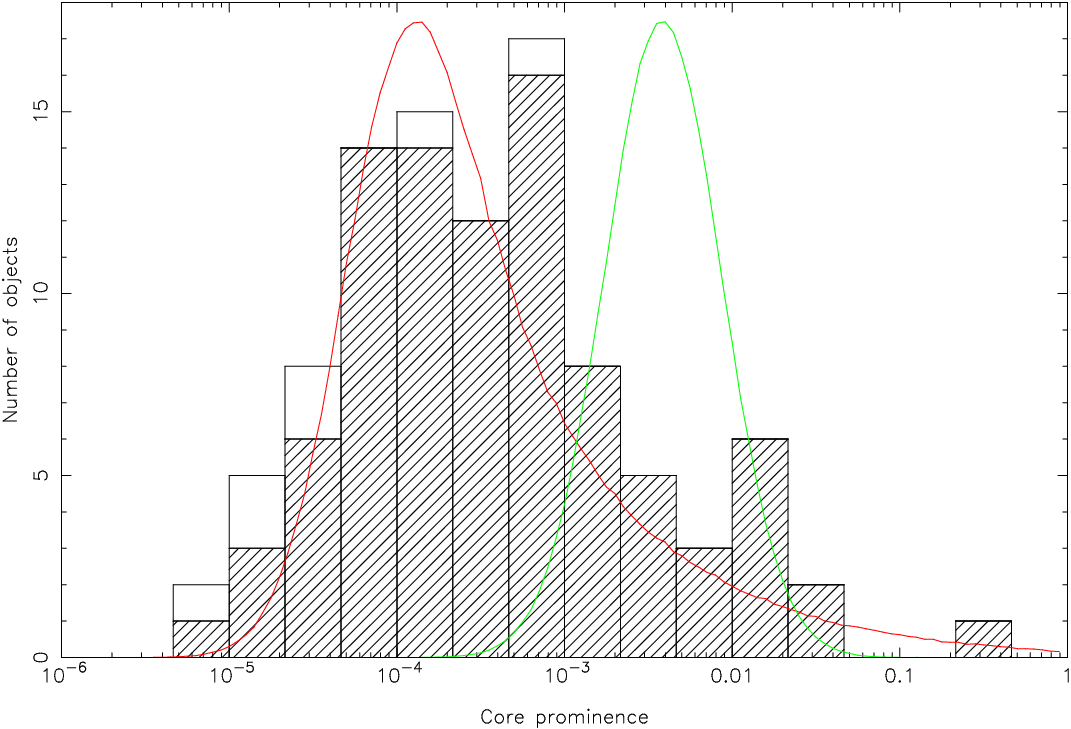}
\caption{\label{fig:coremodel}Histogram of the observed core prominence
  data, $p_{\rm obs_{c}}$. Regions and curves are as in
  Fig.\ \ref{fig:jetmodel}. Here the simulated data have $\gamma = 10.0$
  and $p_{\rm int_{c}}$ has a lognormal distribution with $\sigma=0.8$ (in
  units of the natural logarithm) around $\ln(p_{\rm int_{j}}) = -5.5 $.}
\end{figure}

Examples of simulated data sets using this approach and the beaming
models of equations (\ref{eq:jet_model}) and (\ref{eq:core_model}) are
plotted together with the observed data in Figs \ref{fig:jetmodel} and
\ref{fig:coremodel}. It can be seen that the shape of the resulting
distributions can give a good representation of the true observed
distributions. (Note that the simulated data sets shown are not fitted
to the data in any way other than by simple scaling, though the
parameters used are representative of those we obtain in subsequent
sections by fitting.) The observed jet
and core prominence is strongly determined by $\gamma$ in each case.
The effect of beaming on the shape of the observed prominence with
respect to the normally distributed intrinsic prominence is to
suppress the observed emission in many sources, while boosting it in a
smaller fraction. As $\gamma$ increases this effect is more
pronounced. Changes in the mean of the intrinsic prominence
distribution affect the observed prominence distribution's location on
the $x$-axis but not its shape: it acts as a scale factor. Changes in
the intrinsic scatter that we assume broaden the distribution (and are
thus to some extent degenerate with changes in $\gamma$) but also tend
to smooth out the resulting distribution.

Therefore, while the problem is degenerate, with at least three unknown
parameters (intrinsic prominence, orientation and $\gamma$), it is
clear that analysis of the observed prominence distribution can tell
us something about beaming in the sample.

\begin{table*}
\centering
\begin{minipage}{14cm}
\caption{Glossary for symbols used.}\label{tab:gloss}
\begin{tabular}{lll} \hline
Symbol & Parameter & Definition \\\hline
$\beta$       & speed as a fraction of the speed of light                  & Section \ref{sec:history}\\
$p_{\rm int}$ & intrinsic prominence of the jet or core feature & Section \ref{sec:history}\\
& & \\
$S_{178}$     & total source flux measured at 178 MHz       & Section \ref{sec:sample}\\
& & \\
$F_{\rm obs}$ & measured jet flux                           & Section \ref{sec:jetdef}\\
$B_{\rm obs}$ & jet background flux correction              & Section \ref{sec:jetdef}\\
$J_{\rm obs}$ & background-corrected jet flux               & Section \ref{sec:jetdef}\\
 & & \\
$\alpha$      & spectral index			            & Section \ref{sec:promdef}\\
$P$           & jet or core observed luminosity             & Section \ref{sec:promdef}\\
$R$           & proper distance                             & Section \ref{sec:promdef}\\
$S$           & jet or core flux density                    & Section \ref{sec:promdef}\\
$\alpha_{\rm lf}$ & low frequency spectral index            & Section \ref{sec:promdef}\\
$P_{178}$     & source luminosity, as measured at 178 MHz   & Section \ref{sec:promdef}\\ 
& & \\
$S_{\nu_{\rm obs}}$ & observed flux density                 & Section \ref{sec:beamingmod}\\
$S_{\nu_{\rm rest}}$ & flux density in emitter's rest frame & Section \ref{sec:beamingmod}\\
$\gamma$      & bulk Lorentz factor                         & Section \ref{sec:beamingmod}\\
$m$           & constant reflecting the geometry of the beamed component & Section \ref{sec:beamingmod}\\
& & \\
$p_{\rm obs_{j}}$   & observed jet prominence               & Section \ref{sec:jetmodels} \\
$p_{\rm int_{j}}$  & intrinsic jet prominence               & Section \ref{sec:jetmodels} \\       
$\theta$      & angle of emitter's trajectory with respect to observer's line of sight & Section \ref{sec:jetmodels}\\
& & \\
$p_{\rm obs_{c}}$   & observed core prominence              & Section \ref{sec:coremodels} \\
$p_{\rm int_{c}}$   & intrinsic core prominence              & Section \ref{sec:coremodels} \\
$p_{\rm int_{a}}$  & intrinsic prominence of approaching parsec scale jet & Section \ref{sec:coremodels} \\
$p_{\rm int_{r}}$  & intrinsic prominence of receding parsec scale jet   & Section \ref{sec:coremodels} \\
$\gamma_{\rm a}$ & bulk Lorentz factor of approaching parsec scale jet & Section \ref{sec:coremodels} \\
$\gamma_{\rm r}$ & bulk Lorentz factor of receding parsec scale jet   & Section \ref{sec:coremodels} \\
$\beta_{\rm a}$  & approaching parsec scale jet speed as fraction of speed of light & Section \ref{sec:coremodels} \\
$\beta_{\rm r}$  & receding parsec scale jet speed as fraction of speed of light & Section \ref{sec:coremodels} \\
$\theta_{\rm a}$      & angle of approaching emitter's trajectory with respect to observer's line of sight & Section \ref{sec:coremodels}\\
$\theta_{\rm r}$      & angle of receding emitter's trajectory with respect to observer's line of sight & Section \ref{sec:coremodels}\\
& & \\
$H$             & hypothesis & Section \ref{sec:bayesinf}\\
$D$             & observed data & Section \ref{sec:bayesinf}\\
$I$             & prior information about $D$ & Section \ref{sec:bayesinf}\\
$P(H|D,I)$      & posterior probability of $H$ given $D$ and $I$& Section \ref{sec:bayesinf}\\
$P(H|I)$        & prior probabiliy of $H$ & Section \ref{sec:bayesinf}\\
$P(D|H,I)$      & likelihood of $D$ given $H$ and $I$ & Section \ref{sec:bayesinf}\\
$P(D|I)$        & evidence & Section \ref{sec:bayesinf}\\
$p_{\rm mod}$   & expected observed prominence corresponding to model parameter values & Section \ref{sec:bayesinf}\\
$P(p_{\rm mod_{t+1}}|p_{\rm mod_{t}})$ & transition probability & Section \ref{sec:bayesinf}\\
$p_{\rm can}$   & candidate value for next chain step & Section \ref{sec:bayesinf}\\
$Q(p_{\rm can}|p_{\rm mod_{t}})$ & proposal distribution & Section \ref{sec:bayesinf}\\ 
$\alpha(p_{\rm mod_{t}},p_{\rm can})$ & acceptance probability & Section \ref{sec:bayesinf}\\
$r$             & metropolis ratio & Section \ref{sec:bayesinf}\\
$\lambda$       & factor by which the posterior is scaled in burn-in & Section \ref{sec:bayesinf}\\ 
& & \\
$\bar \gamma$      & mean bulk Lorentz factor & Section \ref{sec:priors}\\
$\gamma_{\rm min}$ & upper limit on $\gamma$ & Section \ref{sec:priors}\\
$\gamma_{\rm max}$ & upper limit on $\gamma$ & Section \ref{sec:priors}\\
$\bar p_{\rm int}$ & mean intrinisic prominence & Section \ref{sec:priors}\\
$\sigma_{\gamma}$ & intrinsic dispersion in $\gamma/\bar \gamma$ & Section \ref{sec:priors}\\
$\sigma_{p_{\rm int}}$ & intrinsic dispersion in prominence & Section \ref{sec:priors}\\
$a$& power-law index for power-law Lorentz factor distribution& Section \ref{sec:priors}\\
& & \\
$X$             & general symbol for parameters & Section \ref{sec:Bayesfac}\\
$M$             & model & Section \ref{sec:Bayesfac}\\
$O_{ik}$        & Bayes factor& Section \ref{sec:Bayesfac}\\ 
& & \\
$\bar \gamma_{\rm j}$ & mean bulk Lorentz factor for the jets &
Section \ref{sec:lumdep}\\
$\bar \gamma_{\rm c}$ & mean bulk Lorentz factor for the cores &
Section \ref{sec:coreres}\\
\hline
\end{tabular}\\
\end{minipage}
\end{table*}

%
%
%
\subsubsection{MCMC approach}\label{sec:bayesinf}
The discussion in the following sections is only a very brief
introduction to the MCMC approach to Bayesian inference; more
detailed treatments can be found in the literature [e.g., Gregory
(2005) and references therein].

We can obtain the posterior probability of a hypothesis $H$ from Bayes' theorem:
\begin{equation}
P(H | D,I) = P(H | I) \frac{P(D|H,I)}{P(D|I)}
\label{eq:bayes}
\end{equation}
where $D$ is the observed data and $I$ is the prior information we have about $D$, $P(H | I)$ is the prior probability of $H$, $P(D|H,I)$ is its likelihood and $P(D|I)$ is the normalization factor, the `evidence'. In the present problem, we have observed prominence data, $p_{\rm obs}$. If we define a parameter space in terms of $p_{\rm int}$, $\gamma$ and $\theta$, we can evaluate the expected observed prominence, $p_{{\rm mod}_{i}}$, corresponding to a parameter set drawn from this space using equation (\ref{eq:jet_model}) or (\ref{eq:core_model}) as appropriate. We can then determine the posterior probability of $p_{{\rm mod}_{i}}$, $P(p_{{\rm mod}_{i}} | p_{\rm obs},I)$, by evaluating $P(p_{{\rm mod}_{i}} | I)$, $P(p_{\rm obs}|p_{{\rm mod}_{i}},I)$ and $P(p_{\rm obs}|I)$.

The obvious problem is that, with a minimum of three parameters, the
exhaustive computation of the probability distribution corresponding
to the defined parameter space is not feasible. Straightforward Monte
Carlo could provide a good approximation -- the method is to draw
uniform, randomly distributed, independent samples from the distribution and its accuracy is determined by the number of these samples. However, this approach is still computationally expensive and much time can be spent in regions where the probability is very small. Instead, MCMC is more efficient as it exploits the fact that samples need {\it not} be drawn independently if they are generated from the target distribution or some function of it, here $P(p_{\rm mod}|p_{\rm obs},I)$, in the correct proportions. 

The Metropolis-Hastings algorithm is used to apply the method. This
algorithm generates a sample set by constructing a walk through
parameter space in which the probability of a sample's being in some
region of space is proportional to the posterior density for that
region. It does this by determining the next chain step, which in this
case is $p_{{\rm mod}_{t+1}}$, with respect to its probability given
the current chain step, $p_{{\rm mod}_{t}}$, through the evaluation of
the transition probability (or kernel), $P(p_{{\rm mod}_{t+1}}|p_{{\rm
mod}_{t}})$.

The algorithm chooses a candidate value for $p_{{\rm mod}_{t+1}}$,
$p_{\rm can}$, from a proposal distribution, $Q(p_{\rm can}|p_{{\rm
mod}_{t}})$ that is understood and easy to evaluate (see Section
\ref{sec:code}). $p_{\rm can}$ is accepted or rejected as $p_{{\rm
mod}_{t+1}}$ as determined by the acceptance probability,
$\alpha(p_{{\rm mod}_{t}},p_{\rm can})$ which can be expressed as
\begin{eqnarray}
\label{eq:metr}
\alpha(p_{{\rm mod}_{t}},p_{\rm can}) & = & {\rm min}(1,r)\\\nonumber & = & {\rm min} \left( 1, \frac{P(p_{\rm can}|D,I)}{P(p_{{\rm mod}_{t}}|D,I)}\frac{Q(p_{{\rm mod}_{t}}|p_{\rm can})}{Q(p_{\rm can}|p_{{\rm mod}_{t}})} \right)
\end{eqnarray}
where $r$ is the Metropolis ratio. If $r \ge 1$, then $p_{\rm can}$ is accepted and $p_{{\rm mod}_{t+1}}=p_{\rm can}$. If $r<1$, then a random variable $U$ is sampled from a uniform distribution in the interval 0 to 1. In the case that $U \le r$, $p_{{\rm mod}_{t+1}}=p_{\rm can}$, otherwise $p_{\rm can}$ is rejected. The transition kernel, the probability that the algorithm will draw and accept a sample $p_{{\rm mod}_{t+1}}$ given the chain's present state, is then
\begin{equation}
P(p_{{\rm mod}_{t+1}}|p_{{\rm mod}_{t}}) = Q(p_{{\rm mod}_{t+1}}|p_{{\rm mod}_{t}})\alpha(p_{{\rm mod}_{t}},p_{\rm can})
\label{eq:trans}
\end{equation}
As a first consideration, we might want the proposal distribution,
$Q(p_{\rm can}|p_{{\rm mod}_{t}})$, to be the target distribution
itself -- but of course, this is unknown as it is this that we are
trying to evaluate. However, if the Markov chain is irreducible,
aperiodic and positive recurrent then it can be shown that there exists a stationary distribution from which all samples will be drawn subsequently once one initial sample is drawn; thus the algorithm will converge on this stationary distribution for a wide range of proposal distributions. 

The probability of drawing $p_{{\rm mod}_{t}}$ from the posterior is $P(p_{{\rm mod}_{t}}|D,I)$ and the probability that $p_{{\rm mod}_{t+1}}$ is subsequently drawn and accepted is given by the joint probability of $p_{{\rm mod}_{t}}$ and $p_{{\rm mod}_{t+1}}$: $P(p_{{\rm mod}_{t}},p_{{\rm mod}_{t+1}})$, which can be expressed as the following:
\begin{equation}
P(p_{{\rm mod}_{t}},p_{{\rm mod}_{t+1}})=P(p_{{\rm mod}_{t}}|D,I) P(p_{{\rm mod}_{t+1}}|p_{{\rm mod}_{t}})
\end{equation}
Expanding this out using equations (\ref{eq:metr}) and (\ref{eq:trans}), gives the detailed balance equation:
\begin{equation}
P(p_{{\rm mod}_{t}}|D,I) P(p_{{\rm mod}_{t+1}}|p_{{\rm mod}_{t}}) = P(p_{{\rm mod}_{t+1}}|D,I) P(p_{{\rm mod}_{t}}|p_{{\rm mod}_{t+1}})
\end{equation}
From this it can be seen that the stationary distribution is the target distribution of the chain, irrespective of the proposal distribution -- $Q(p_{\rm can}|p_{{\rm mod}_{t}})$ -- initially used. The sampling process before the stationary distribution is reached is referred to as burn-in.

The algorithm allows the chain to move to regions of increasing
probability while sometimes accepting a chain step of lower
probability. This contributes to the efficient exploration of
parameter space as the chain can move away from regions of local
maxima. However, for a target posterior distribution of a
multi-dimensional problem this flexibility will not necessarily be
sufficient to prevent the chain becoming stuck in a local maximum
region; the results will still be dependent on the starting sample. In
practice we use multiple chains with information exchange (see the
next section) but we also make use of a simulated annealing method
that allows a modified posterior distribution to be sampled during burn-in.

This modified posterior distribution is defined as 
\begin{equation}
p(p_{\rm obs}|D,I) = p(p_{\rm obs}|I)p(D|p_{\rm obs},I)^{\lambda}
\label{eq:lambda}
\end{equation}
where $\lambda$ may take values between 0 and 1. When $\lambda=0$ the
samples are drawn just from the prior distribution, which will
typically be much flatter than the likelihood function -- if this is
the case increasing $\lambda$ draws samples from an increasingly
peaked function until the posterior itself is being sampled when
$\lambda=1$. Sampling from the flatter distributions gives the sampler
better opportunity to reach all regions of the posterior probability
distribution, even in the presence of many local maxima. The general
practice is to determine the rate at which $\lambda$ is increased from
0 to 1 (the annealing schedule) such that it will correspond to the
burn-in period. In this scheme, once $\lambda=1$ sampling is being
made from the target distribution. The rate at which $\lambda$ should
be increased is then subjective and determined by experiment.
%
%

\subsubsection{Implementation}
\label{sec:code}

Our implementation of the Metropolis-Hastings algorithm is heavily
based on the approach of Hobson \& Baldwin (2004), as implemented in
the {\sc metro} sampler code, kindly provided to us by Mike Hobson.
We implemented the basic algorithm in modular C using the Message
Passing Interface (MPI) framework to allow it to run on a cluster of
multi-core computers; only the functions that implement the likelihood
function and priors need be modified for a particular problem. At
run-time the code separates into one master and one or more slave
threads; in general it is advantageous to have as many slave threads
as there are available CPUs.

Each slave carries
out a separate and (except as discussed below) independent
Metropolis-Hastings run with burn-in (a `chain'), starting in a random
point in the prior parameter space. As discussed by Hobson \& Baldwin,
multi-threading of this kind greatly reduces the chances of getting
stuck in a local minimum in parameter space; it also allows us to
speed up by a factor of the number of available computing cores the sampling of
the posterior probability distribution. We borrow from the {\sc metro}
code the concept of `engines', which are different versions of the
proposal distribution $Q$. Several engines are available, and the code
picks the one to use for each trial at random, using probabilities
that we have assigned based on our experience with the code. The
active engines in the implementation of the code used here are `take a
random step in a single dimension of the problem', `take a random jump
in all dimensions of the problem simultaneously', `jump to a random
point in parameter space' and `jump in one dimension to the current
position of another thread' (allowing cross-mixing). The sizes of the
random steps used in the first two engines are adaptively chosen
during burn-in to achieve a reasonable ratio of success to failure, but
are fixed thereafter.

The master thread records the accepted samples
of all the slave threads in a file, and also co-ordinates
synchronization and communication between the slave threads. The final
result when all slave threads are completed is a large file giving the
co-ordinates in parameter space of all accepted samples both before
and after burn-in. The density of points in a given (necessarily
finite) region of $n$-dimensional parameter space is proportional to
the posterior probability of the model parameters lying in that
regions. Estimated value determination, credible interval
estimates, plotting and evidence determination can then all be carried
out using this file. The burn-in points are discarded for most
applications, but are used for evidence determination, as discussed below.

\subsubsection{Model parameters and priors}\label{sec:priors}

Our basic model for jets and cores (hereafter `the basic model')
has up to 4 parameters.

\begin{enumerate}
\item $\bar \gamma$, the mean bulk Lorentz factor. We adopt a uniform
  (uninformative) prior in the range 1..$\gamma_{\rm max}$ for this
  quantity. $\gamma_{\rm max}$ throughout our fitting is taken to be
  5.5 for jets and 20.0 for cores.
\item $\bar p_{\rm int}$, the mean intrinsic prominence. As this is a scale
  parameter, we adopt a uniform prior in $\ln(\bar p_{\rm int})$ between two
  values chosen to cover all reasonable values of parameter space.
  This avoids bias towards large values (Gregory 2005).
\item $\sigma_\gamma$, the intrinsic dispersion in the bulk Lorentz
  factor. Lorentz factors of simulated sources are drawn from a normal
  distribution with mean $\bar\gamma$ and standard deviation
  $\bar\gamma\sigma_\gamma$, truncated so that $\gamma \ge 1$ in all cases. The
  case $\sigma_\gamma = 0$ is equivalent to a delta function in
  Lorentz factor. We adopt a uniform prior for $\sigma_\gamma$ between
  0 and 0.6.
\item $\sigma_{p_{\rm int}}$, the intrinsic dispersion in the prominence. Since
  the observed prominence will be the product of a number of
  independent variables, it is appropriate to draw simulated 
  prominences from a lognormal distribution with mean
  $\ln(\bar p_{\rm int})$ and standard deviation $\sigma_{p_{\rm int}}$; the case
  $\sigma_{p_{\rm int}} = 0$ is equivalent to a delta function in intrinsic
  prominence. We adopt a uniform prior for $\sigma_{p_{\rm int}}$ between 0 and
  6.
\end{enumerate}

For cores, we also investigated a model (hereafter `the power-law
  model', as discussed in Section \ref{sec:modelval}) in which the Lorentz factor
follows a power-law distribution between two limits, $\gamma_{\rm
  min}$ and $\gamma_{\rm max}$, with a power-law index $a$, i.e.
$P(\gamma) \propto \gamma^{-a}$. For these parameters we adopted
uniform priors between 1 and 5 for $\gamma_{\rm
  min}$, between 10 and 40 for $\gamma_{\rm max}$, and between 0 and 5
for $p$. Since the mean intrinsic prominence $\bar p_{\rm int}$ is a
required component of this model, and a dispersion in the intrinsic
prominence, parametrized as above by $\sigma_{p_{\rm int}}$, may also
be considered, the model has up to 5 parameters.
%
\subsubsection{Likelihood calculation}\label{sec:like}
The likelihood is the probability of obtaining a given set of intrinsic prominences, $p_{\rm obs}$, given
the model and the priors, i.e, $\prod_k P(p_{{\rm obs}_{k}}|p_{\rm
  mod},I)$. In general it is difficult to write down the likelihood
function for the type of models discussed in Section \ref{sec:priors}
analytically. We therefore proceed by Monte Carlo methods. For the set
of model parameters determined by the Metropolis-Hastings algorithm,
we simulate a set of $N$ model prominences based on the beaming equations (equations
\ref{eq:jet_model} and \ref{eq:core_model}) and the dispersions
discussed in Section \ref{sec:priors} and with an appropriate
distribution of the angle to the line of sight $\theta$ (in practice
we assume sources randomly oriented to the line of sight in all
models, so $\theta$ values are drawn from a uniform distribution in $\cos\theta$). By construction, these
simulated prominences are distributed with the appropriate probability
distribution for the model being tested. The probability of obtaining
any given observed value of the prominence, $p_{{\rm obs}_{k}}$, with an
associated error, assumed Gaussian, of $\sigma_{p_{{\rm obs}_{k}}}$
given the model is
then given by
\begin{equation}
P(p_{{\rm obs}_{k}}|p_{\rm mod},I) = {1\over N} \sum^{N}_{i=1}  \exp \left \{ - \frac{ (p_{{\rm obs}_{k}}-p_{{\rm mod}_{i}})^{2}}{2 \sigma^{2}_{p_{{\rm obs}_{k}}}} \right \}
\label{eq:like}
\end{equation}
Essentially here we are Monte Carlo integrating over
the product of the probability distribution for the data point and the
probability distribution for the given model, with a suitable normalization.
In the case where $p_{\rm obs}$ is an upper limit, we could simply write
\begin{equation}
P(p_{{\rm obs}_{k}}|p_{\rm mod},I) = {1\over N} \sum^{N}_{i=1} l(p_{{\rm
  obs}_{k}},p_{{\rm mod}_{i}})
\end{equation}
where
\begin{equation}
l(a,b) = \begin{cases}1,&b\le a\\
                0,&{\rm otherwise}\\
\end{cases}
\end{equation}
In practice we slightly `soften' the treatment of limits in the case
where $p_{{\rm obs}_{k}} < p_{{\rm mod}_{i}}$ to take into account that no
limit is absolute; thus model data points with $p_{{\rm mod}_{i}} > p_{{\rm
obs}_{k}}$ are assigned a non-zero probability, again based on a normal
distribution, so as to treat the limits as though they were $3\sigma$
upper limits. This approach is exactly valid for the limits on core
prominences and an adequate approximation for the less well-defined
jet limits.

A two- to five- dimensional parameter space is defined by the possible
parameters given in Section \ref{sec:priors}. The procedure described
above effectively integrates over $\theta$ and the posterior
probability represents a joint posterior probability for some or all
of $\bar p_{\rm int}$, $\sigma_{p_{\rm int}}$, $\bar \gamma$,
$\sigma_{\gamma}$, $\gamma_{\rm min}$, $\gamma_{\rm max}$ or $a$, depending on how the parameter space has been defined. 
We are more interested in the
posterior probability of some of these parameters than of others: in
particular the one of most physical interest is $\gamma$. 
The {\it marginal} posterior probability of $\gamma$, or in general of
any other parameter, can be determined from the joint posterior
probability by integrating over the other parameters:
\begin{equation}
P(\bar \gamma|p_{\rm obs},M)= \int dX P(\bar \gamma, X|p_{\rm obs}, M)
\end{equation}
where $X$ represents all parameters except $\bar \gamma$. This
integration can be carried out trivially using the output of the MCMC routine.

\subsubsection{The evidence and the Bayes factor}\label{sec:Bayesfac}
The denominator of equation (\ref{eq:bayes}) is the `evidence' and is calculated from the following.
\begin{equation}
P(p_{\rm obs}|I)=\sum_{i}P(p_{{\rm mod}_{i}}|I)P(p_{\rm obs}|p_{{\rm mod}_{i}},I)
\label{eq:evid}
\end{equation}
Since this is the same for each $p_{\rm mod_{i}}$ for a given model,
$M$, it may be omitted in calculating the posterior probability of $p_{\rm mod_{i}}$ (equation \ref{eq:bayes}).
However, if we want to compare models, for example a model in which we
have 2 parameters, $\bar \gamma$ and $\bar p_{\rm int}$, with one where we have
four, $\bar \gamma$, $\bar p_{\rm int}$, $\sigma_{\gamma}$ and $\sigma_{p_{\rm int}}$,
the evidence must be evaluated. This is not possible using only the
post burn-in samples from the posterior, as these are, by definition, not
uniformly sampled over the prior. Instead, it is possible to use the burn-in samples themselves, as justified by the following.

The continuous equivalent of equation (\ref{eq:evid}) gives us the global likelihood for $M$, that is the weighted average likelihood for its parameters.
\begin{equation}
P(p_{\rm obs}|M)=\int dX P(X|I)P(p_{\rm obs}|X,M)
\label{eq:glike}
\end{equation}
where $X$ represents all parameters.

Remembering that the function actually being sampled by the code is a modification of $P(X|p_{\rm obs},I)$, (equation \ref{eq:lambda}), we can define a partition function, $Z(\lambda)$ as
\begin{align}
Z(\lambda) & = \int dX P(X|M,I)P(p_{\rm obs}|X,M,I)^{\lambda}\\
& = \int dX \exp\{\ln[P(X|M,I)]+\lambda \ln[P(p_{\rm obs}|X,M,I)]\}
\label{eq:partfunction}
\end{align}
$Z(\lambda)$ is then the tempering simulation corresponding to $\lambda$, and integrating over all $\lambda$ gives us the function we want, the global likelihood,
\begin{equation}
P(p_{\rm obs}|M) = \int^{1}_{0} d \ln Z(\lambda) 
\end{equation}
But, from equation (\ref{eq:partfunction}) with some rearrangement, we can write the derivative of $\ln[Z(\lambda)]$ as
\begin{equation}
\frac{d}{d\lambda}\ln[Z(\lambda)] = \langle \ln[P(p_{\rm obs}|M,X,I)] \rangle_{\lambda}
\end{equation}
where $\langle \ln[P(p_{\rm obs}|M,X,I)] \rangle_{\lambda}$ is the expectation value of $\ln[P(p_{\rm obs}|M,X,I)]$ and the subscript $\lambda$ denotes which tempering simulation the samples correspond to. But we can also say that
\begin{align}
\ln[P(p_{\rm obs}|X,M,I) & = \int \lambda \langle \ln[P(p_{\rm obs}|X,M,I)] \rangle_{\lambda}\\
& \approx \frac{1}{i}\sum_{i} \lambda_{i} \langle \ln[P(p_{\rm obs}|X,M,I)] \rangle_{\lambda_{i}}
\end{align}
So for a given model, the log global likelihood can be obtained from
the burn-in samples and this, along with the model prior, can be used
to evaluate the odds ratio, defined as the ratio of the evidence
values for the two models:
\begin{equation}
O_{1,2} = \frac{P(M_{1}|I)}{P(M_{2}|I)} \frac{P(p_{\rm obs}|M_{1},I)}{P(p_{\rm obs}|M_{2},I)}
\end{equation}
In this analysis identical priors are applied to the various models
tested, so this expression is simplified to the ratio of global
likelihoods, known as the Bayes factor. We can then use the odds ratio
or, equivalently, the Bayes factor, to attempt to say which of two
models provides the best description of the data.
\subsubsection{Credible intervals and regions}
\label{sec:cred}

We define the credible interval on a parameter for a confidence level
$p$ as the smallest interval such that the posterior probability of
the parameter lying in the interval is $p$. In one dimension (i.e.
integrating over all other parameters, which in the context of the
output of the code simply means ignoring their values) there is
clearly, provided the posterior probability has a single peak, a
unique choice of interval that satisfies this requirement, which can
be found by an exhaustive search over the results of the sampler, and
which is accurate up to the constraints imposed by the finite sampling
of the posterior. In more than one dimension, there is no such obvious
choice, and we find an approximation to the credible region by binning
the posterior probability distribution (marginalizing out
uninteresting parameters), sorting by the probability of
each binned element, and taking the first $m$ elements that sum to
give the probability $p$: the credible region is then approximately
the region of parameter space enclosing these $m$ grid elements. This
procedure has the advantage that if there truly are multiple peaks in
the posterior they will be represented correctly: it has the
disadvantage that the results may depend on the binning and, in
particular, that the binning may have to be quite coarse in order to
define a credible region in many dimensions. In what follows we only
present credible intervals in one dimension and credible regions in
two dimensions.

We note in passing that this definition of the credible interval can
(for a very asymmetrical posterior) actually exclude the position of
the Bayesian estimate of the parameter, since it corresponds to the
mode, while the Bayesian estimate corresponds to the mean; the
credible interval is certainly not constrained to lie symmetrically
about the mean (in fact it is more likely to be symmetrical about the
maximum-likelihood value). This should be borne in mind when interpreting the
`errors' that we quote on certain parameters.

%
\subsection{Method verification}
%
\label{sec:testing}
\subsubsection{Data simulation}
Before running the code on real datasets it is necessary to establish
whether the Bayesian estimator of the parameters of interest
provided by the code (i.e. the mean of the values of that parameter
for all post burn-in samples) is really an unbiased estimate of the true value
under realistic conditions and whether the uncertainties (credible
intervals) on parameters or combinations of parameters are good
estimates of the true uncertainty.

This can be done using Monte Carlo simulations (as H99 did for their
K-S test model fitting). To do this we simulated a large number (50)
of sets of jet data, with parameters chosen to represent a best guess
at what resembles the real data most closely. A simulated dataset
contained 100 prominence points, each with $p_{\rm obs_{j}}$ simulated using
equation (\ref{eq:jet_model}). 

Initially $p_{\rm int_{j}}$ was drawn from a Gaussian distribution
with $\ln(\bar p_{\rm int})=-7,\sigma_{p_{\rm int}} = 0.6$ and we used
$\bar\gamma=1.5, \sigma_\gamma = 0$. For 30 per cent of sources
$p_{\rm obs_{j}}$ was not used as calculated but instead an upper
limit was generated. To generate the upper limits we used a
  random number between 0.1 and the calculated jet prominence value.
  While the non-detection of a jet might be expected to be dependent
  on $p_{\rm obs_{j}}$ to some extent, when we considered the jet
  prominences and upper limits for the sample as a whole (Paper I) we
  found that there was no discernible trend with prominence with
  respect to upper limits. It would appear that observational effects,
  particularly the variations in observing resolution, mask any strong
  dependence of jet detection on $p_{\rm obs_{j}}$ such that there is
  no obvious model that should be used for generating upper limits;
  hence we adopt a random deviate method. It is reasonable to expect
  that this method, along with the high maximum upper limit relative
  to the observed upper limit values used, will only provide simulated
  data of `worse' quality than that which is observed, justifying the
  simplifying approach in this context. For the remaining sources,
the jet prominence was used with an associated error. This error was
drawn from a normal distribution where the mean was the calculated
$p_{\rm obs_{j}}$ and the standard deviation was variable, itself
taken from a normal distribution with (0,$0.1\times p_{\rm j}$). This
standard deviation was chosen to simulate as closely as possible the
error distribution of the real jet data.
\subsubsection{Parameter estimation}\label{sec:paraerr}
For each simulated dataset we then ran the MCMC code and derived the
posterior probability distribution, from which we could obtain the
Bayesian estimators of the value of each parameter. The mean and
dispersion of the values of $\langle \bar\gamma \rangle$, $\langle
\ln(\bar p_{\rm int}) \rangle$ and $\langle \sigma_{p_{\rm int}}
\rangle$ for each run of the code -- where the angle brackets denote
the Bayesian estimates of the model parameters -- then tell us whether
the code recovers the true values and estimates the uncertainties
appropriately. For the initial trial input values of $\bar\gamma$ and
$\ln(\bar p)$ we obtained a mean $\langle \bar\gamma \rangle$ of 1.52
and a standard deviation of 0.14, a mean $\langle \ln(\bar p_{\rm
  int}) \rangle$ of -7.01 with standard deviation 0.16 and a mean
$\langle \sigma_{p_{\rm int}} \rangle$ of 0.56 with standard deviation
0.16. We can see that while the errors and limits imposed on the data
result in considerable scatter in the returned parameters about the
true (input) values, the means are in good agreement with those true
values once the scatter is taken into account. Thus there is no
evidence that the code is biased for this type of dataset. Moreover,
the standard deviations we find (which are in some sense frequentist
estimates of the uncertainties on the fitting procedure, and were used
in that way by H99) are very similar in magnitude to the credible
intervals inferred from the posterior probability distributions for
individual simulations. We verified that the same result holds for a
range of other values of $\bar\gamma$. The fitting procedure is,
unsurprisingly, biased if we fit the wrong model (e.g., if we fit a
simulated dataset with intrinsic dispersion with a model in which
$\sigma_{p_{\rm int}}$ is held at zero, then we will recover an
artificially high value of $\bar\gamma$) but for models that match the
data it is not. This is true for all parameters of the basic
  model and also for the parameters of the power-law model for cores
  when we carried out appropriate simulations, although we noted in
  simulating power-law distributions that it is very hard to recover
  good constraints on $\gamma_{\rm max}$, which is only constrained by
  the tail of the prominence distribution.

\subsubsection{Model selection}\label{sec:modsel}
We further investigated whether the Bayesian evidence could reliably
be used to distinguish between models. Again concentrating on
simulated datasets with $\sigma_\gamma$ set to zero and with
properties matched to the observed jet data, we simulated a large
number of datasets with $\sigma_{p_{\rm int}} > 0$ and investigated
whether the Bayes factor allowed us to distinguish between models in
which (a) the fitted $\sigma_{p_{\rm int}}$ was fixed to zero and (b)
it was allowed to vary. The results turned out to depend on the value
of $\sigma_{p_{\rm int}}$ adopted in the simulated data, as one would
expect. For very low values of the simulated $\sigma_{p_{\rm int}}$,
the Bayes factor on average favoured the simpler model, even though it
is formally incorrect: the data here simply do not give us enough
information to require the extra parameter. For larger values of the
input $\sigma_{p_{\rm int}}$, $\ga 0.2$, the Bayes factor clearly
tends to favour the models with non-zero intrinsic dispersion.
However, when we fitted the same data with other versions of the basic
model with intrinsic dispersion (e.g. models where $\sigma_{p_{\rm
    int}}$ was constrained to be zero and $\sigma_\gamma$ was allowed
to vary) we found that the Bayes factor did not reliably distinguish
between the correct model and other possible models for the
dispersion; all such models gave very similar Bayes factors.
Consistent with this, we found that the Bayes factor in simulations
did not allow us to distinguish between versions of the basic model
with Gaussian dispersion either in the intrinsic prominence or in the
Lorentz factor and the power-law model for cores; again, the Bayes
factor is sufficent to point to some dispersion but not to say what
its origin is.

\subsubsection{Observed data}
Finally we note that two factors distinguish the jet and core data.
The core data contains far fewer upper limits than the jet data and so
in one sense is expected to be easier to fit. However, as is evident
from the broader distribution of core prominences (Fig.\
\ref{fig:coremodel}) and also expected from earlier work (e.g. H99) we
expect significantly larger $\bar\gamma$ values in the fits to the
cores. This means that the feature of the data that discriminates
between different models becomes the extent of the tail of very high
prominences. However, since there are only a very few sources with
such high values, the fit becomes less reliable. In a fit
to 50 simulated jet data sets with input $\bar\gamma = 5.0$ and
$\sigma_{p_{\rm int}} = 0.6$, we found a mean fitted $\bar\gamma$ of 5.3 with
standard deviation 1.0. The higher standard deviation of the fit
results to the simulated data as a fraction of the true $\gamma$, when
compared to the jet data, indicates that the second factor dominates:
even though the core data are in some sense better, the fitting
problem is harder and the results likely to be more uncertain.

We conclude that the MCMC fitting procedure can give a good estimate
of the true underlying beaming parameters in the presence of the types
of error and the numbers of upper limits seen in the real data. In the
next section we proceed to use this procedure to estimate parameters
using the actual measurements.

%
%
\section{Results}\label{sec:results}
%
%
\subsection{Jets}
\label{sec:jets}

Since the spectral indices of jets are poorly known, we carried
  out all the modelling in this section for two representative values
  of jet spectral index, $\alpha = 0.5$ and $\alpha = 0.8$
  (recalculating the jet prominence distribution consistently for the
  $\alpha = 0.8$ case from the original data of Paper I.) By using
  these two values, which are close to the extreme values observed for
  individual objects, we can both evaluate the extent to which a
  choice of $\alpha$ affects our results and, hopefully, bracket the
  true values of the beaming parameters for intermediate spectral
  index values.

\begin{figure*}
\includegraphics[angle=0,width=8.5cm]{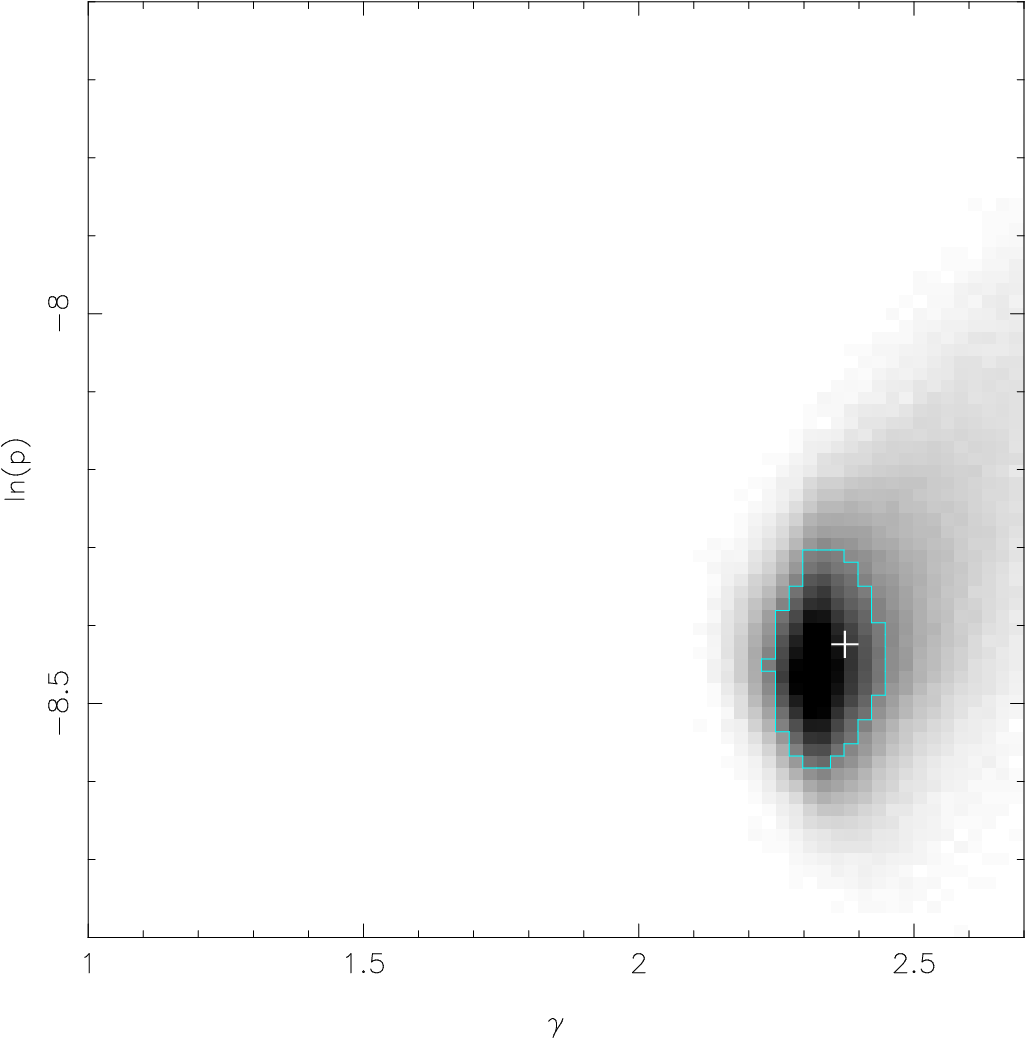}
\includegraphics[angle=0,width=8.5cm]{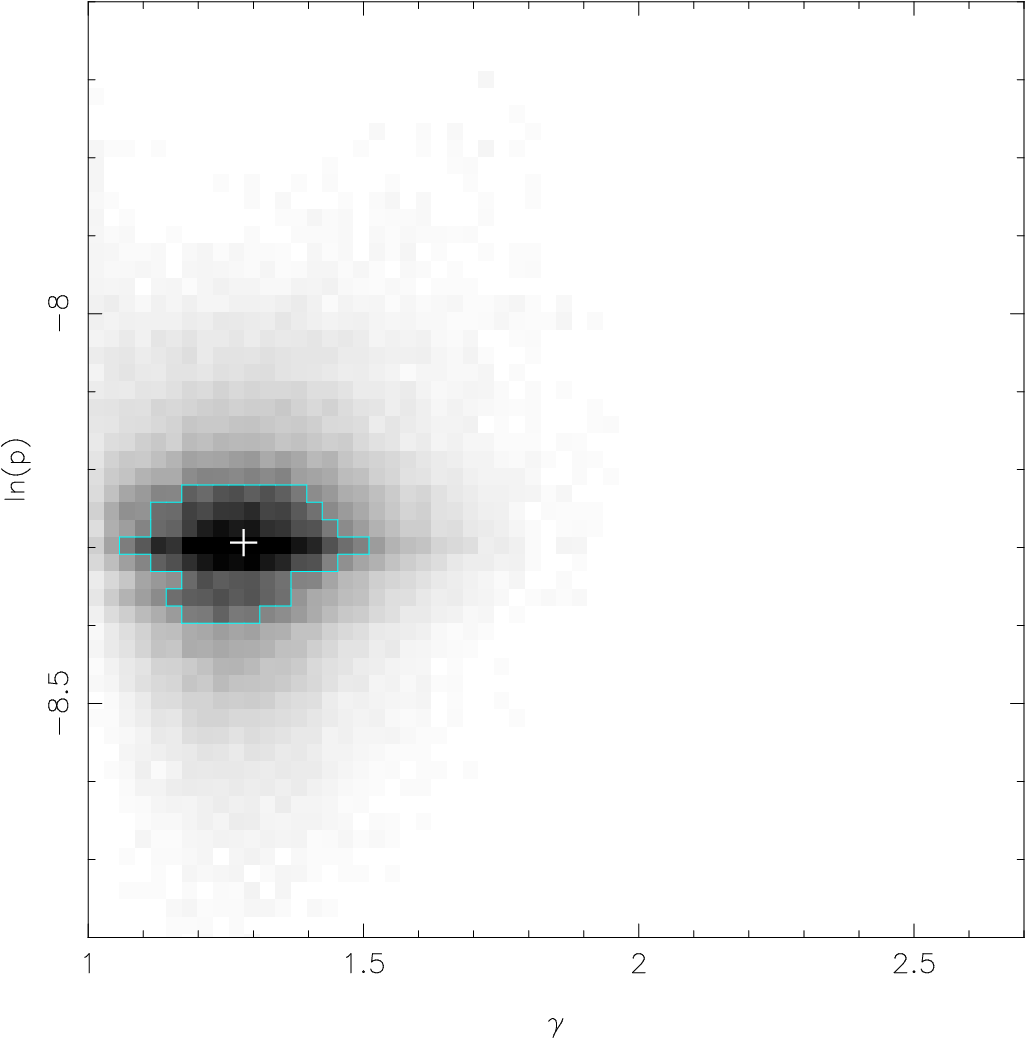}
\includegraphics[angle=0,width=8.5cm]{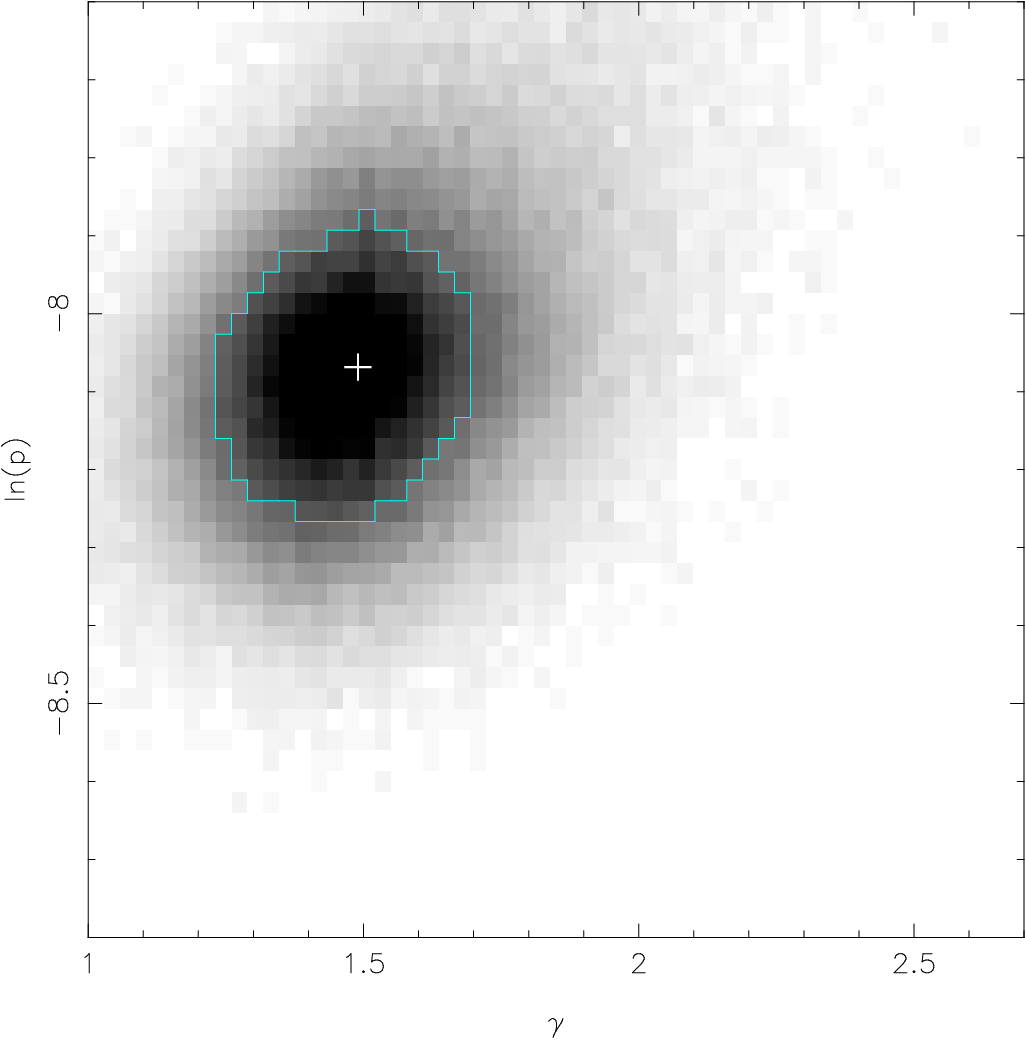}
\includegraphics[angle=0,width=8.5cm]{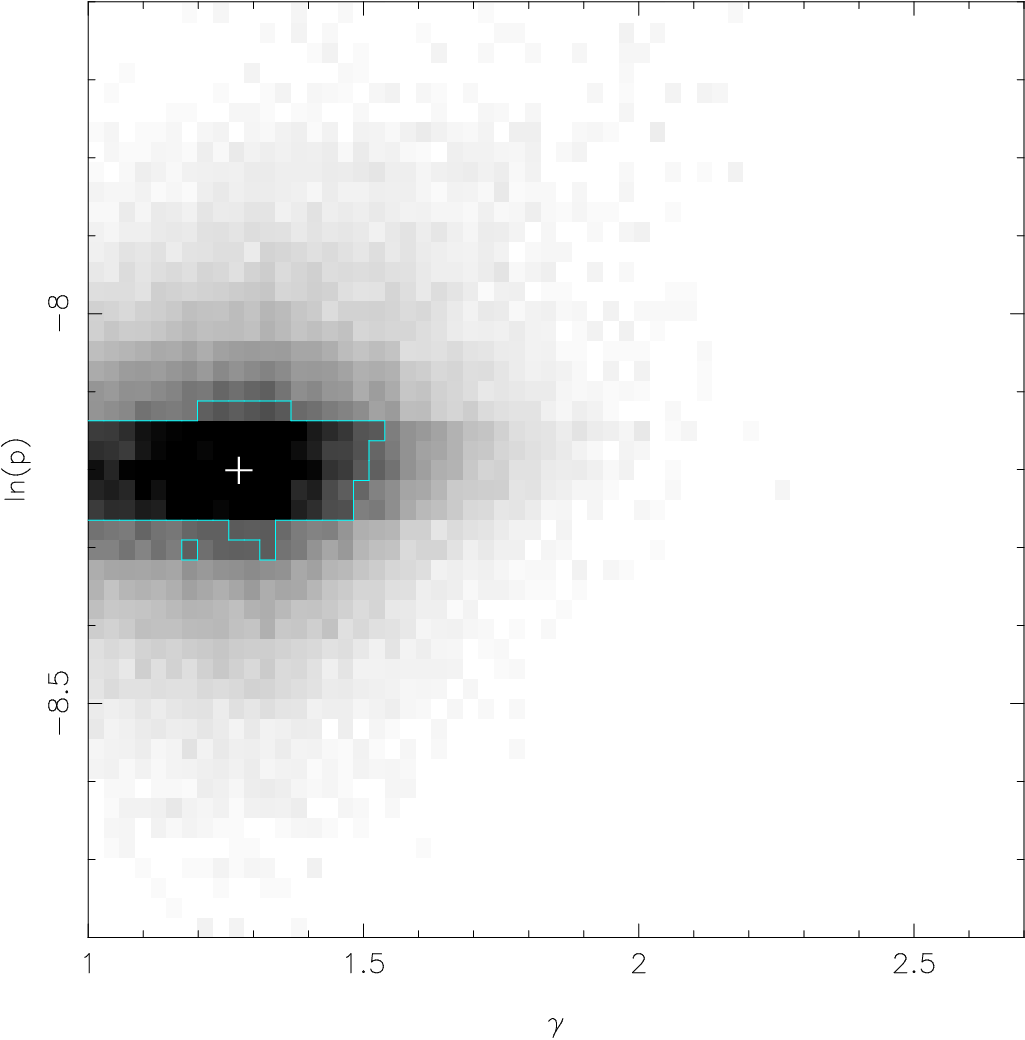}
\caption{The posterior probability distribution for fits to the jet
  prominence distribution for the whole sample. Contours show the
  smallest region that contains 68 per cent of the posterior
  probability distribution, i.e. the two-dimensional Bayeian credible
  region as defined in Section \ref{sec:cred}. The white cross marks the position of the Bayesian estimator
  of $p_{\rm int}$ and $\gamma$. Top left: a delta-function distribution is
  assumed for both the intrinsic prominence and the Lorentz factor
  (i.e. $\sigma_\gamma = 0$, $\sigma_{p_{\rm int}} = 0$). Top right:
  $\sigma_\gamma = 0$ but $\sigma_{p_{\rm int}}$ is allowed to vary. Bottom
  left: $\sigma_{p_{\rm int}}=0$ but $\sigma_\gamma$ is allowed to vary. Bottom
  right: both $\sigma_{p_{\rm int}}$ and $\sigma_\gamma$ are allowed to vary.
  All plots are marginalized over $\sigma_{p_{\rm int}}$ and $\sigma_\gamma$
  for ease of comparison.}
\label{fig:jet-pp}
\end{figure*}

We initially used the simplest possible model, with an assumed
delta-function distribution in both the intrinsic prominence and the
bulk Lorentz factor, to estimate beaming parameters from the prominence distribution of the jets for
all 98 objects in the sample. We then
allowed the parameters governing the intrinsic dispersion in
prominence and Lorentz factor to vary. Results are
shown in Fig.\ \ref{fig:jet-pp} (for the $\alpha = 0.5$ case) and
tabulated in Table \ref{tab:jet-results}.
(It should be noted that the posterior distribution of $\sigma_\gamma$
is clearly limited by the prior: the upper limit on the prior
distribution was chosen so as to avoid generating large numbers of
negative Lorentz factors.)

Comparison of the values for the different $\alpha$ values in
  Table \ref{tab:jet-results} shows that, as expected, the choice of jet
spectral index has a significant effect on the result. Although the
two sets of results are very comparable, the results for the higher
spectral index have systematically lower estimated Lorentz factors.
The next thing to note is that the Lorentz factor for a
delta-function distribution in both parameters is quite high:
$\bar\gamma = 2.37$ corresponds to $\beta = 0.91$, and $\bar\gamma =
2.07$ to $\beta = 0.88$, which can be
compared to the $\beta = 0.62$ found by H99. However, when we allow
any dispersion in either the intrinsic prominence or the Lorentz
factor, the values of $\bar\gamma$ drop very substantially.
$\bar\gamma = 1.28$ corresponds to $\beta = 0.62$ and $1.18$ to $\beta
= 0.53$: H99 found $\beta
\approx 0.5$ for a similar model. We emphasise that no model allows
jet speeds with $\gamma \sim 10$; this would predict a much larger
dispersion than is seen in the data. The ratios of evidence (Bayes
factors) very strongly favour any model with intrinsic dispersion, but
does not allow us to distinguish between the three different models
that have this property with any degree of confidence (as we would
expect from the simulations discussed above). In fact, when we fitted
a model with $\bar\gamma$ fixed to 1.0 and $\sigma_\gamma = 0$ (i.e.
the data are modelled with {\it only} a lognormal prominence
distribution) the Bayes factor marginally favours this simplest model
over any other, showing that the prominence data in themselves provide
no evidence for a beaming model, while intrinsic scatter in the data
is required.

Since there is some evidence that the low-excitation radio galaxies
(LERGs) do not participate in the unified models for narrow-line radio
galaxies, broad-line radio galaxies and quasars (see H99 for details)
we also fitted the same set of models to the prominence data excluding
the 15 low-excitation objects in the sample. The very similar results
we obtained are tabulated in Table \ref{tab:jet-results}. As the
inclusion or exclusion of the small number of LERGs clearly makes
little difference to the results, we do not show plots of the
posterior probability distribution.

\begin{table*}
\caption{Results of fits to the jet prominence distribution. Fits are
  carried out for two values of the assumed jet spectral index, $\alpha$. The Bayes
  factor quoted is the natural log of the ratio between the evidence
  value for the $\sigma_\gamma = 0$, $\sigma_{p_{\rm int}} = 0$ fit
  (for a given sample and $\alpha$) and the
  others; thus it gives a measure of the degree to which the other
  fits are preferred. Errors quoted are 68 per cent credible intervals
  marginalizing over all other parameters. Where no errors are quoted,
  the quantity concerned was fixed at the stated level.}
\label{tab:jet-results}
\begin{tabular}{lrlrrrrr}
\hline
Sample&$\alpha$&Model&$\bar\gamma$&$\ln(\bar p_{\rm int})$&$\sigma_{p_{\rm int}}$&$\sigma_\gamma$&Bayes
factor\\
\hline
Jets, all sources&$0.5$&$\sigma_\gamma = 0$, $\sigma_{p_{\rm int}} = 0$&
$2.37^{+0.03}_{-0.11}$&$-8.42^{+0.06}_{-0.11}$&0&0&--\\
&&   $\sigma_\gamma=0$, $\sigma_{p_{\rm int}}$ free&
$1.28^{+0.10}_{-0.11}$&$-8.29^{+0.06}_{-0.05}$&$0.95^{+0.10}_{-0.09}$&0&19.4\\
&&   $\sigma_\gamma$ free, $\sigma_{p_{\rm int}} = 0$&
$1.49^{+0.12}_{-0.17}$&$-8.07^{+0.12}_{-0.13}$&0&$0.46^{+0.12}_{-0.05}$&20.0\\
&&   $\sigma_\gamma$ free, $\sigma_{p_{\rm int}}$ free&
$1.27^{+0.13}_{-0.18}$&$-8.20^{+0.06}_{-0.06}$&$0.76^{+0.15}_{-0.13}$&$0.30^{+0.18}_{-0.15}$&19.5\\
&&   $\bar\gamma = 1.0$, $\sigma_\gamma=0$, $\sigma_{p_{\rm int}}$ free&
1.0&$-7.91^{+0.11}_{-0.10}$&$1.18^{+0.05}_{-0.06}$&0&21.0\\
\hline
&$0.8$&$\sigma_\gamma = 0$, $\sigma_{p_{\rm int}} = 0$&
$2.07^{+0.02}_{-0.06}$&$-8.50^{+0.05}_{-0.09}$&0&0&--\\
&&   $\sigma_\gamma=0$, $\sigma_{p_{\rm int}}$ free&
$1.18^{+0.05}_{-0.07}$&$-8.36^{+0.08}_{-0.03}$&$0.94^{+0.08}_{-0.07}$&0&19.2\\
&&   $\sigma_\gamma$ free, $\sigma_{p_{\rm int}} = 0$&
$1.24^{+0.09}_{-0.15}$&$-8.18^{+0.10}_{-0.10}$&0&$0.45^{+0.11}_{-0.06}$&20.2\\
&&   $\sigma_\gamma$ free, $\sigma_{p_{\rm int}}$ free&
$1.16^{+0.04}_{-0.16}$&$-8.25^{+0.06}_{-0.07}$&$0.64^{+0.19}_{-0.15}$&$0.32^{+0.15}_{-0.13}$&19.3\\
&&   $\bar\gamma = 1.0$, $\sigma_\gamma=0$, $\sigma_{p_{\rm int}}$ free&
1.0&$-7.90^{+0.11}_{-0.09}$&$1.11^{+0.04}_{-0.06}$&0&21.0\\
\hline
Jets, LERGs excluded&$0.5$&$\sigma_\gamma = 0$, $\sigma_{p_{\rm int}} = 0$&
$2.43^{+0.04}_{-0.14}$&$-8.42^{+0.07}_{-0.13}$&0&0&--\\
&&   $\sigma_\gamma=0$, $\sigma_{p_{\rm int}}$ free&
$1.32^{+0.11}_{-0.12}$&$-8.35^{+0.07}_{-0.06}$&$0.94^{+0.09}_{-0.11}$&0&14.4\\
&&   $\sigma_\gamma$ free, $\sigma_{p_{\rm int}} = 0$&
$1.51^{+0.14}_{-0.19}$&$-8.12^{+0.12}_{-0.14}$&0&$0.46^{+0.14}_{-0.04}$&15.6\\
&&   $\sigma_\gamma$ free, $\sigma_{p_{\rm int}}$ free&
$1.31^{+0.15}_{-0.19}$&$-8.24^{+0.06}_{-0.07}$&$0.73^{+0.18}_{-0.13}$&$0.32^{+0.19}_{-0.14}$&14.5\\
&&   $\bar\gamma = 1.0$, $\sigma_\gamma=0$, $\sigma_{p_{\rm int}}$ free&
1.0&$-7.97^{+0.07}_{-0.05}$&$1.19^{+0.05}_{-0.05}$&0&15.8\\
\hline
&$0.8$&$\sigma_\gamma = 0$, $\sigma_{p_{\rm int}} = 0$&
$2.11^{+0.02}_{-0.08}$&$-8.50^{+0.06}_{-0.11}$&0&0&--\\
&&   $\sigma_\gamma=0$, $\sigma_{p_{\rm int}}$ free&
$1.21^{+0.06}_{-0.10}$&$-8.38^{+0.06}_{-0.06}$&0&$0.97^{+0.11}_{-0.07}$&15.6\\
&&   $\sigma_\gamma$ free, $\sigma_{p_{\rm int}} = 0$&
$1.30^{+0.12}_{-0.15}$&$-8.20^{+0.13}_{-0.11}$&0&$0.45^{+0.13}_{-0.06}$&15.4\\
&&   $\sigma_\gamma$ free, $\sigma_{p_{\rm int}}$ free&
$1.20^{+0.09}_{-0.16}$&$-8.26^{+0.04}_{-0.06}$&$0.61^{+0.24}_{-0.17}$&$0.34^{+0.16}_{-0.12}$&14.0\\
&&   $\bar\gamma = 1.0$, $\sigma_\gamma=0$, $\sigma_{p_{\rm int}}$ free&
1.0&$-7.94^{+0.05}_{-0.06}$&$1.15^{+0.03}_{-0.06}$&0&15.5\\
\hline
\end{tabular}
\end{table*}

\subsection{Luminosity dependence in jets}\label{sec:lumdep}

As mentioned above (Section \ref{sec:jets}), our best estimates of
$\gamma$ for jet models either with or without intrinsic dispersion
tend to lie above the corresponding values found by H99. To some
extent this is expected, since (i) we take limits and errors into
account properly, which H99's method did not permit, and a correct
treatment of the limits at least would be expected to broaden the
effective distribution, and (ii) we are quoting not the
maximum-likelihood value but the Bayesian estimator of the Lorentz
factor, $\int\gamma p(\gamma) {\rm d}\gamma$, which will be biased
towards higher values with respect to the maximum-likelihood estimator
if $p(\gamma)$ has a long tail to higher values, as it undoubtedly
does in some fits (Fig.\ \ref{fig:jet-pp}). In addition, H99 used a
different definition of prominence, normalizing with respect to the
total high-frequency flux, which might have tended to reduce the scatter in
the observed prominence if any of the high-frequency structure (e.g.
hotspots) had been affected by beaming.

However, it is worth asking whether any component of this difference
could correspond to a real physical difference between our sample and
that of H99. The two samples overlap to some extent -- of the 31 NLRG
studied by H99, 22 are also in our sample -- but the main difference
is that H99 considered only objects with $z<0.3$, and hence largely
with low luminosities. We began our investigation by fitting our
models with $\sigma_{p_{\rm int}} = 0$ and $\sigma_\gamma = 0$ to the
jet data of H99, for consistency treating upper limits as detections
and assuming a constant fractional error on each prominence
measurement. For consistency with the convention adopted by H99, we use $\alpha = 0.8$
  throughout this comparison. We found a Bayesian estimate of $\bar\gamma =
1.65^{+0.01}_{-0.07}$ ($\beta = 0.80$), which is considerably larger
than the best-fitting value found by H99, $\beta = 0.62$. This
suggests that there is some difference in the different fitting
approaches, and perhaps reinforces the notion that H99's approach of
fitting using the K-S statistic was not ideal. If we allow
$\sigma_{p_{\rm int}}$ to vary in the fits, which was the most
sophisticated model considered by H99, we obtain $\bar\gamma =
1.10^{+0.02}_{-0.07}$; this corresponds to $\beta = 0.43$, which is
very similar to the $\beta \approx 0.5$ reported by H99.

Although this is at best qualified success in reproducing the results
of H99, we noted that the values of $\gamma$ for the two models we
fitted lay systematically below the results we obtained for fitting
the same models to our sample as a whole (Section \ref{sec:jets},
Table \ref{tab:jet-results}). This motivated us to look for luminosity
effects in the jet speeds in our sample. For simplicity we divided
the whole sample, including LERGs, at a 178-MHz luminosity of $5 \times 10^{26}$ W
Hz$^{-1}$ sr$^{-1}$, which is the cutoff luminosity we used in a
number of statistical tests in Paper I. This gives a sample of 40
low-luminosity and 58 high-luminosity objects. When we fitted the
model with $\sigma_{p_{\rm int}} = 0$ and $\sigma_\gamma = 0$ to these
two datasets (here using $\alpha = 0.5$ only), we found that $\bar\gamma = 2.37^{+0.09}_{-0.28}$ for
the low-luminosity sample and $1.92^{+0.03}_{-0.09}$ for the
high-luminosity one, a marginally significant difference but one that
is in the {\it opposite} sense to that implied by the results above
from the H99 data. When we fitted the model with $\sigma_\gamma = 0$
but $\sigma_{p_{\rm int}}$ free, we found $\bar\gamma =
1.25^{+0.06}_{-0.21}$ for the low-luminosity sample and
$1.55^{+0.12}_{-0.14}$ for the high-luminosity one. However, the
intrinsic dispersion for the low-luminosity sample is higher
($\sigma_{p_{\rm int}} = 1.04^{+0.11}_{-0.12}$ versus $0.65^{+0.13}_{-0.17}$) so it is
plausible that we are simply seeing different tradeoffs in the
somewhat degenerate plane of $\gamma$ versus $\sigma_{p_{\rm int}}$.
If, finally, we fix $\sigma_{p_{\rm int}}$ to its best-fitting value
of 0.95 for the whole sample (Table \ref{tab:jet-results}) and repeat
the fits, the estimates of $\gamma$ become $1.29^{+0.08}_{-0.18}$ and
$1.33^{+0.11}_{-0.16}$ respectively; these are not significantly
different from each other or from the estimate for the sample as a
whole. We conclude that there is no convincing evidence for a trend in
$\bar\gamma_{\rm j}$ with source luminosity.
\subsection{Cores}\label{sec:coreres}
Using the basic model, we find that, for a delta-function
distribution in Lorentz factor and intrinsic prominence, $\bar p_{\rm
  int}$ and $\bar\gamma$ are constrained to lie along a line in
parameter space, as found in previous work (see Hardcastle \etal\ 2003
for the equation of this line) and we can only really say that
$\bar\gamma \ga 10$ (the limit $\bar\gamma < 20$ is imposed by the
prior). However, when any intrinsic dispersion is introduced,
$\bar\gamma$ becomes reasonably well constrained, with Bayesian
estimators in the range 10--14, though again this is affected by the
choice of prior. As in the case of the jets, the Bayes factors for
these models strongly favour some intrinsic dispersion, but we cannot
distinguish between different dispersion models. Models with no
beaming and only intrinsic dispersion are not favoured in this
dataset, presumably because the tail of the prominence distribution is
not well modelled with a lognormal distribution.

Again, we also fitted the same set of models to the subsample that
excludes the LERGs. Here the tendency is to reduce the values of
$\bar\gamma$ required -- perhaps because the LERGs include a
significant number of both core non-detections with strong upper
limits and bright cores with high prominence values. The results of
fitting the basic model to both datasets are tabulated in Table
\ref{tab:core-results}.

We then estimated the parameters of the power-law model for the core
prominence distribution. These results are tabulated in Table
\ref{tab:core-results-pow}. We initially fixed $\sigma_{p_{\rm int}}$
to 0, so that all the dispersion comes from the power-law distribution
of $\gamma$; with this prior we obtain some constraints on the
remaining parameters of the model, though they are still strongly
affected by the choice of prior (particularly $\gamma_{\rm max}$,
which we expect from the simulated data to be unconstrained). We
attempted a fit with a fixed $a$ value of 2.0, as favoured by some
previous work (e.g. Liu \& Zhang 2007) but this is strongly
disfavoured by the Bayes factor. When $\sigma_{p_{\rm int}}$ is free,
the parameters of the power-law model are essentially unconstrained,
although the Bayes factor favours such a model. It can be seen that,
as expected from the simulations, these data give us no reason to
favour a power-law distribution of Lorentz factors over a lognormal
one; the very slight difference in Bayes factors in favour of the
power-law model when $\sigma_{p_{\rm int}}$ is free is not
significant.

\subsection{Luminosity dependence in cores}\label{sec:corelumdep}

In Paper I, we suggested that the statistically significant
  difference in core prominence seen in our low-luminosity and
  high-luminosity samples might be evidence for a dependence of
  Lorentz factor on source luminosity. To investigate this we divided
  the sample of core prominences by luminosity in the same way as was
  done for jets in Section \ref{sec:lumdep}. Fitting only the version
  of the basic model in
  which $\sigma_{p_{\rm int}}$ is free, since, as discussed above,
  only models with some intrinsic dispersion give reasonably
  well-constrained $\bar\gamma$ values, we found $\bar\gamma =
  11.20^{+4.87}_{-5.66}$ for the low-luminosity sources and
  $10.67^{+3.51}_{-6.43}$ for the high-luminosity sources. Thus there
  is no evidence for differences in Lorentz factor, although clearly
  $\bar\gamma$ is poorly constrained. The intrinsic normalization
  values for the low- and high-luminosity sample are respectively
  $-4.70^{+1.26}_{-0.55}$ and $-6.06^{+1.21}_{-0.84}$, while the
  estimates of $\sigma_{p_{\rm int}}$ are $1.40^{+0.15}_{-0.24}$ and
  $1.19^{+0.14}_{-0.23}$. The most obvious interpretation of these
  results is that, if any luminosity dependence is present in our
  data, it is a luminosity dependence of intrinsic prominence rather
  than of bulk Lorentz factor; either explanation is of course equally
  good in terms of reproducing our original observation of a
  difference in observed prominence. Given these results and the fact
  that in any case the data do not appear to allow us to distinguish
  between any but the simplest of models for core Lorentz factor
  distribution, we have not attempted to fit a model in which core
  Lorentz factor depends on source luminosity to the data.

\begin{figure*}
\includegraphics[angle=0,width=8.5cm]{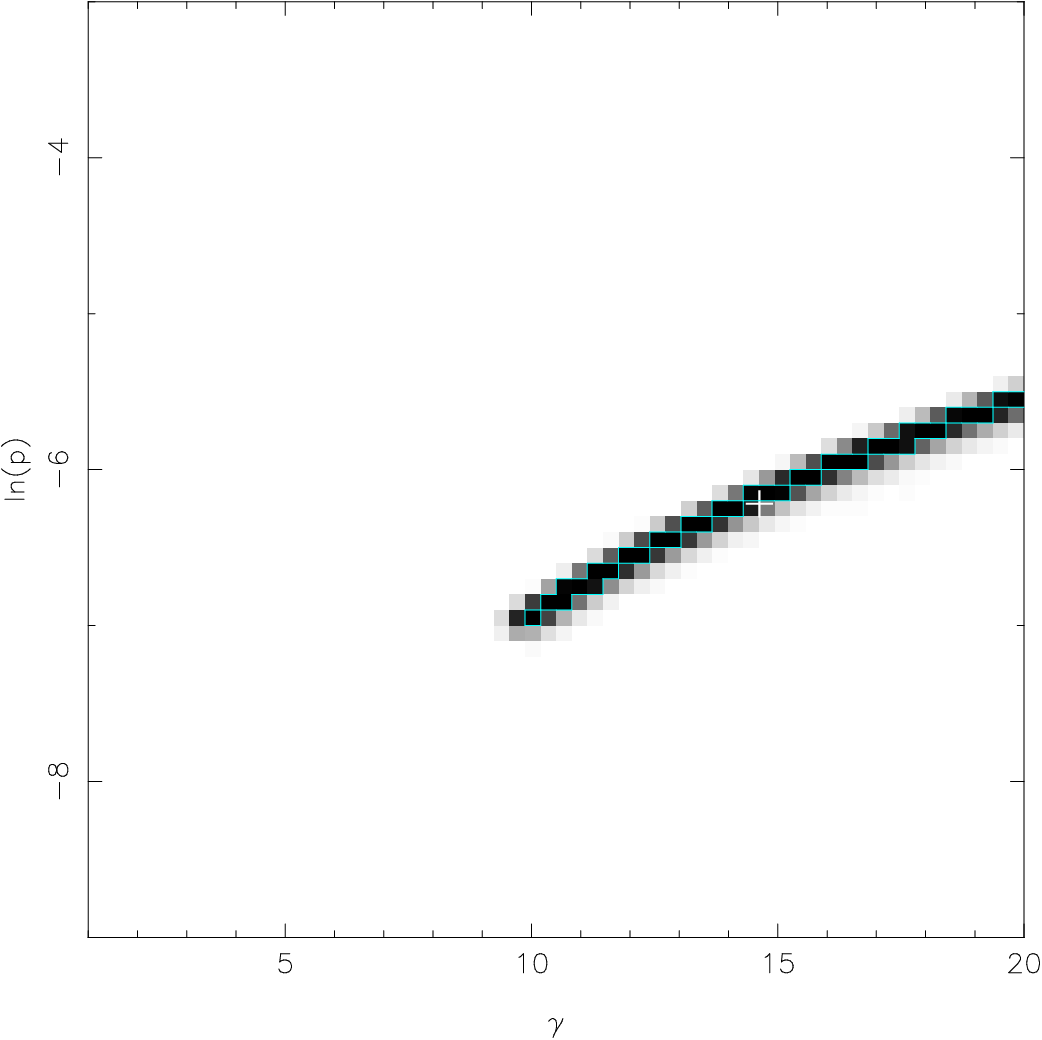}
\includegraphics[angle=0,width=8.5cm]{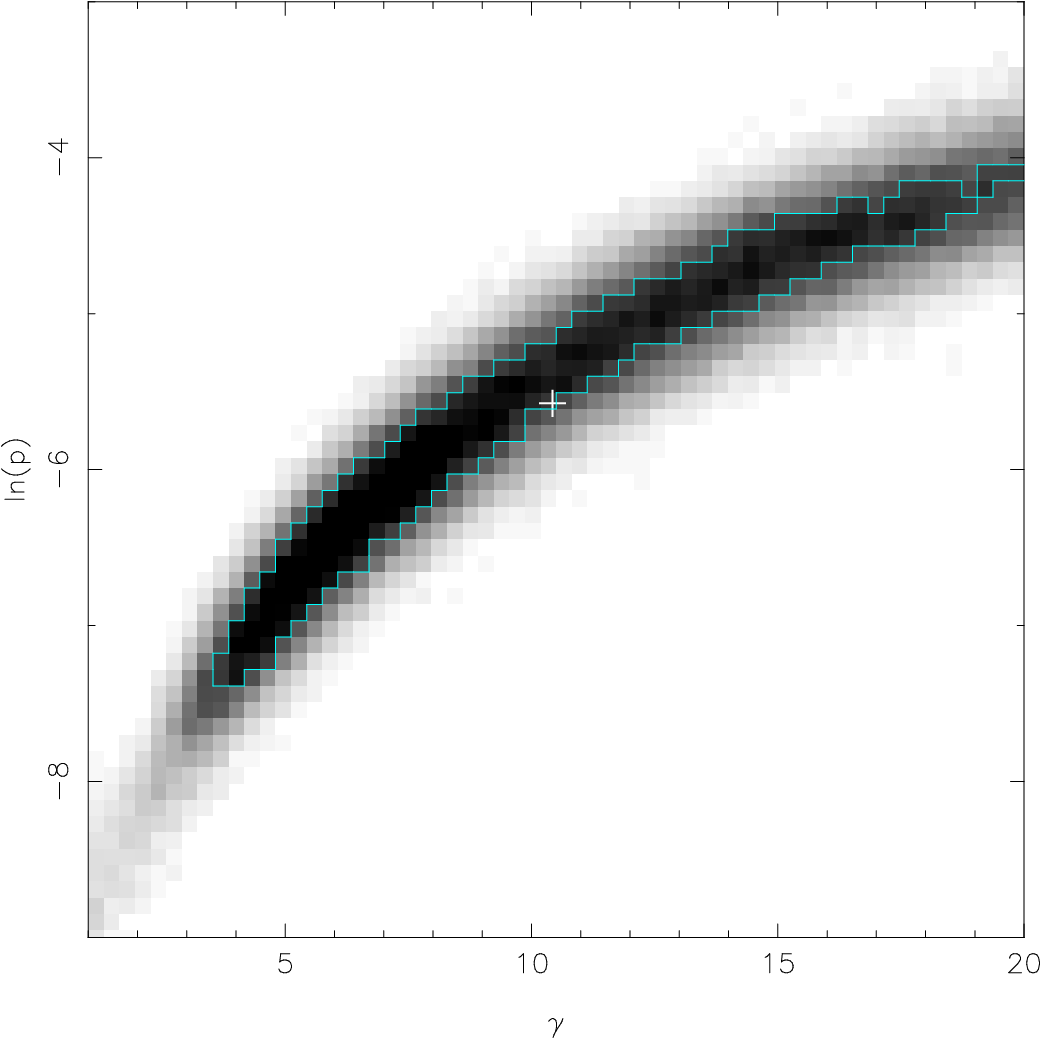}
\includegraphics[angle=0,width=8.5cm]{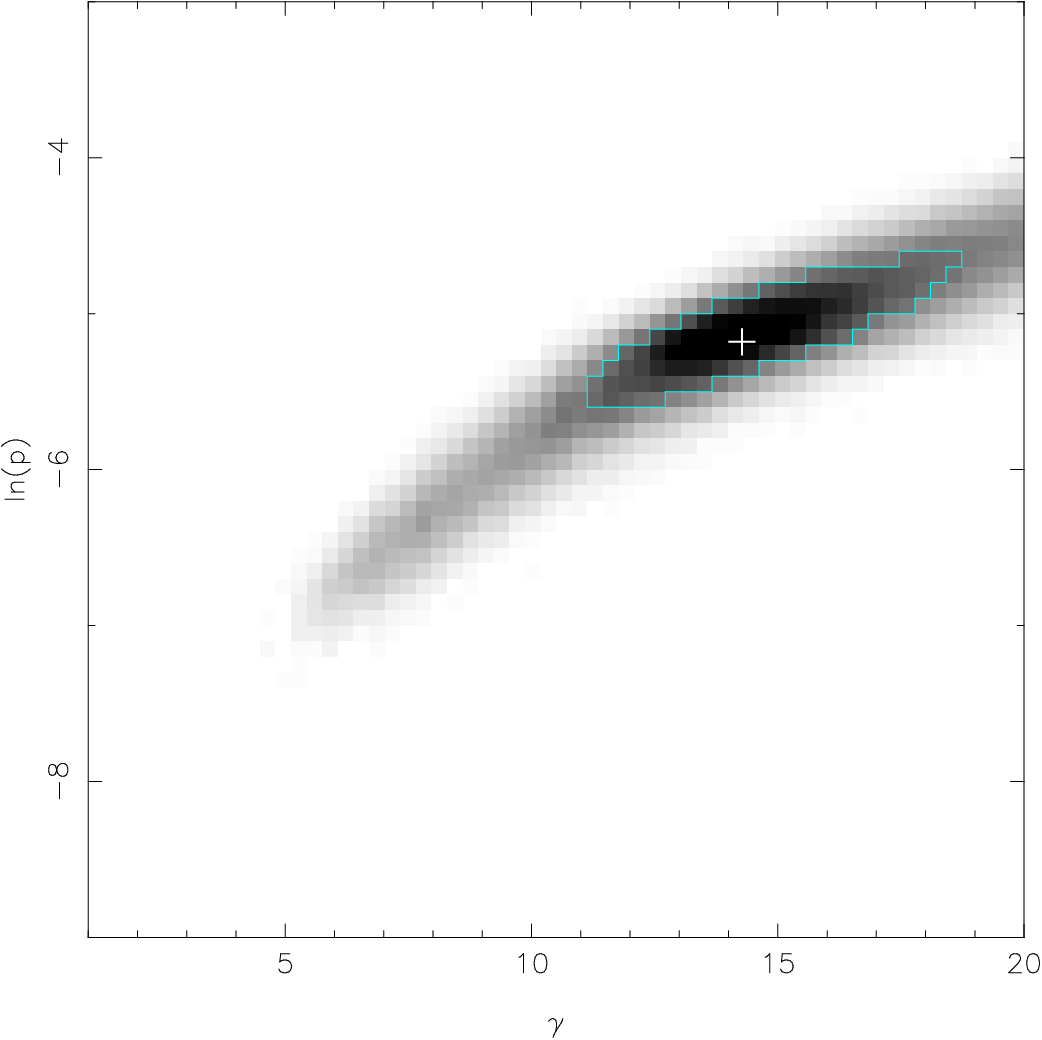}
\includegraphics[angle=0,width=8.5cm]{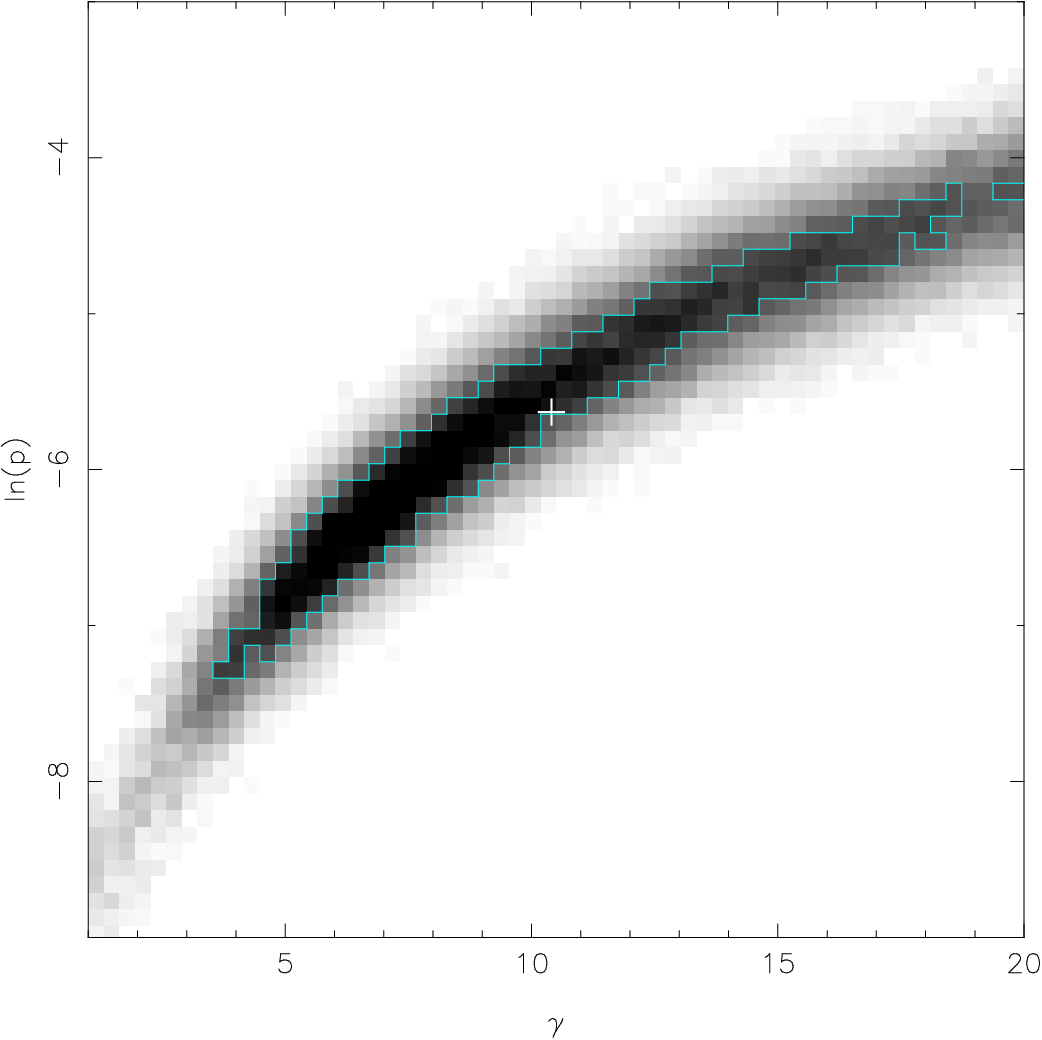}
\caption{The posterior probability distribution for fits to the core
  prominence distribution. Labelling as for Fig.\ \ref{fig:jet-pp}. Top
  left: a delta-function distribution is assumed for both the
  intrinsic prominence and the Lorentz factor (i.e. $\sigma_\gamma =
  0$, $\sigma_{p_{\rm int}} = 0$). Top right: $\sigma_\gamma = 0$ but
  $\sigma_{p_{\rm int}}$ is allowed to vary. Bottom left: $\sigma_{p_{\rm int}}=0$ but
  $\sigma_\gamma$ is allowed to vary. Bottom right: both $\sigma_{p_{\rm int}}$
  and $\sigma_\gamma$ are allowed to vary. All plots are marginalized
  over $\sigma_{p_{\rm int}}$ and $\sigma_\gamma$ for ease of comparison.}
\label{fig:core-pp}
\end{figure*}

\begin{table*}
\caption{Results of fits of the basic model to the core prominence distribution.
  Conventions as for Table \ref{tab:jet-results}.}
\label{tab:core-results}
\begin{tabular}{llrrrrr}
\hline
Sample&Model&$\bar\gamma$&$\ln(\bar p_{\rm int})$&$\sigma_{p_{\rm int}}$&$\sigma_\gamma$&Bayes
factor\\
\hline
Cores, all sources&$\sigma_\gamma = 0$, $\sigma_{p_{\rm int}} = 0$&
$14.62^{+2.02}_{-4.71}$&$-6.22^{+0.65}_{-0.28}$&0&0&--\\
&   $\sigma_\gamma=0$, $\sigma_{p_{\rm int}}$ free&
$10.42^{+3.15}_{-6.54}$&$-5.57^{+1.31}_{-0.72}$&$1.43^{+0.12}_{-0.13}$&0&33.6\\
&   $\sigma_\gamma$ free, $\sigma_{p_{\rm int}} = 0$&
$14.27^{+2.58}_{-2.44}$&$-5.18^{+0.40}_{-0.24}$&0&$0.57^{+0.03}_{-0.01}$&32.9\\
&   $\sigma_\gamma$ free, $\sigma_{p_{\rm int}}$ free&
$10.40^{+3.03}_{-5.91}$&$-5.63^{+1.02}_{-0.84}$&$1.23^{+0.21}_{-0.17}$&$0.33^{+0.27}_{-0.09}$&33.7\\
&   $\bar\gamma = 1.0$, $\sigma_\gamma=0$, $\sigma_{p_{\rm int}}$ free&
1.0&$-8.68^{+0.11}_{-0.12}$&$2.15^{+0.09}_{-0.11}$&0&32.7\\
\hline
Cores, LERGs excluded&$\sigma_\gamma = 0$, $\sigma_{p_{\rm int}} = 0$&
$11.96^{+2.75}_{-6.39}$&$-6.69^{+1.17}_{-0.53}$&0&0&--\\
&   $\sigma_\gamma=0$, $\sigma_{p_{\rm int}}$ free&
$8.24^{+2.35}_{-5.46}$&$-6.28^{+0.96}_{-1.45}$&$1.31^{+0.11}_{-0.16}$&0&27.8\\
&   $\sigma_\gamma$ free, $\sigma_{p_{\rm int}} = 0$&
$12.66^{+5.67}_{-2.82}$&$-5.52^{+0.91}_{-0.28}$&0&$0.55^{+0.05}_{-0.01}$&27.9\\
&   $\sigma_\gamma$ free, $\sigma_{p_{\rm int}}$ free&
$8.84^{+2.49}_{-5.64}$&$-6.13^{+1.03}_{-1.16}$&$1.09^{+0.22}_{-0.21}$&$0.34^{+0.26}_{-0.08}$&28.0\\
&   $\bar\gamma = 1.0$, $\sigma_\gamma=0$, $\sigma_{p_{\rm int}}$ free&
1.0&$-8.80^{+0.11}_{-0.10}$&$2.02^{+0.09}_{-0.08}$&0&27.8\\
\hline

\end{tabular}
\end{table*}

\begin{table*}
\caption{Results of fits of the power-law model to the core prominence
  distribution. Conventions as for Table \ref{tab:jet-results}. The
  Bayes factor is with respect to the corresponding model with
  $\sigma_\gamma = 0$, $\sigma_{p_{\rm int}} = 0$ in Table \ref{tab:core-results}.}
\label{tab:core-results-pow}
\begin{tabular}{llrrrrrr}
\hline
Sample&Model&$\ln(\bar p_{\rm int})$&$\sigma_{p_{\rm
    int}}$&$\gamma_{\rm min}$&$\gamma_{\rm max}$&$a$&Bayes
factor\\
\hline
Cores, all sources&$\sigma_{p_{\rm int}} = 0$, $a$ free
&$-6.10^{+0.34}_{-0.23}$&0&$4.11^{+0.89}_{-0.25}$&$30.74^{+8.16}_{-3.83}$&$0.15^{+0.02}_{-0.15}$&31.1\\
&$\sigma_{p_{\rm int}} =0$, $a=2$
&$-3.12^{+0.76}_{-0.26}$&0&$2.49^{+0.49}_{-1.49}$&$29.46^{+10.54}_{-2.99}$&2&24.0\\
&$\sigma_{p_{\rm int}}$ free, $a$ free&$-6.77^{+0.47}_{-0.35}$&$1.40^{+0.21}_{-0.23}$&$3.77^{+1.17}_{-0.38}$&$24.24^{+5.29}_{-14.13}$&$2.60^{+2.09}_{-1.10}$&35.0\\
\hline
Cores, LERGs excluded&$\sigma_{p_{\rm int}} = 0$, $a$ free&
$-6.74^{+0.29}_{-0.34}$&0&$3.47^{+0.43}_{-0.69}$&$26.56^{+7.02}_{-8.79}$&$0.43^{+0.07}_{-0.42}$&26.9\\
&$\sigma_{p_{\rm int}} =0$, $a=2$&$-7.07^{+0.20}_{-0.26}$&0&$3.90^{+0.41}_{-0.53}$&$28.12^{+11.87}_{-3.95}$&2&24.1\\
&$\sigma_{p_{\rm int}}$ free, $a$ free&$-7.10^{+0.48}_{-0.40}$&$1.21^{+0.25}_{-0.27}$&$3.44^{+0.78}_{-0.89}$&$24.35^{+5.29}_{-14.35}$&$2.63^{+2.31}_{-0.85}$&28.8\\
\hline
\end{tabular}
\end{table*}

%
%
\subsection{Core-jet correlation}

Fits to the jet and core prominence distributions alone do not take
into account the correlation between the two quantities. In a model in
which the prominence distributions were completely dominated by
intrinsic scatter, we would expect no correlation between these two
quantities, but in fact we showed in Paper I that there is such a
correlation which is significant even in the presence of upper limits.
It is straightforward to modify our approach to simulate the
two-dimensional probability distribution of jet and core prominences
and to sum the likelihoods of objects in (jet, core) prominence space.
Accordingly we carried out some fits using this approach, fitting, as
before, both to the full database and to the subset of sources that
excludes the LERGs. Since we have seen that intrinsic dispersion is
required in the fits to the two quantities individually, we included
this in our two-dimensional models; for simplicity we restricted
ourselves to a model in which there are intrinsic dispersions in the
prominences but not in the beaming speeds, so that the free parameters
of the model are $\bar p_{\rm int_c}$, $\sigma_{p_{\rm int_c}}$,
$\bar\gamma_{\rm c}$, $\bar p_{\rm int_j}$, $\sigma_{p_{\rm int_j}}$
and $\bar\gamma_{\rm j}$.

\begin{figure*}
\epsfxsize 8.5cm
\epsfbox{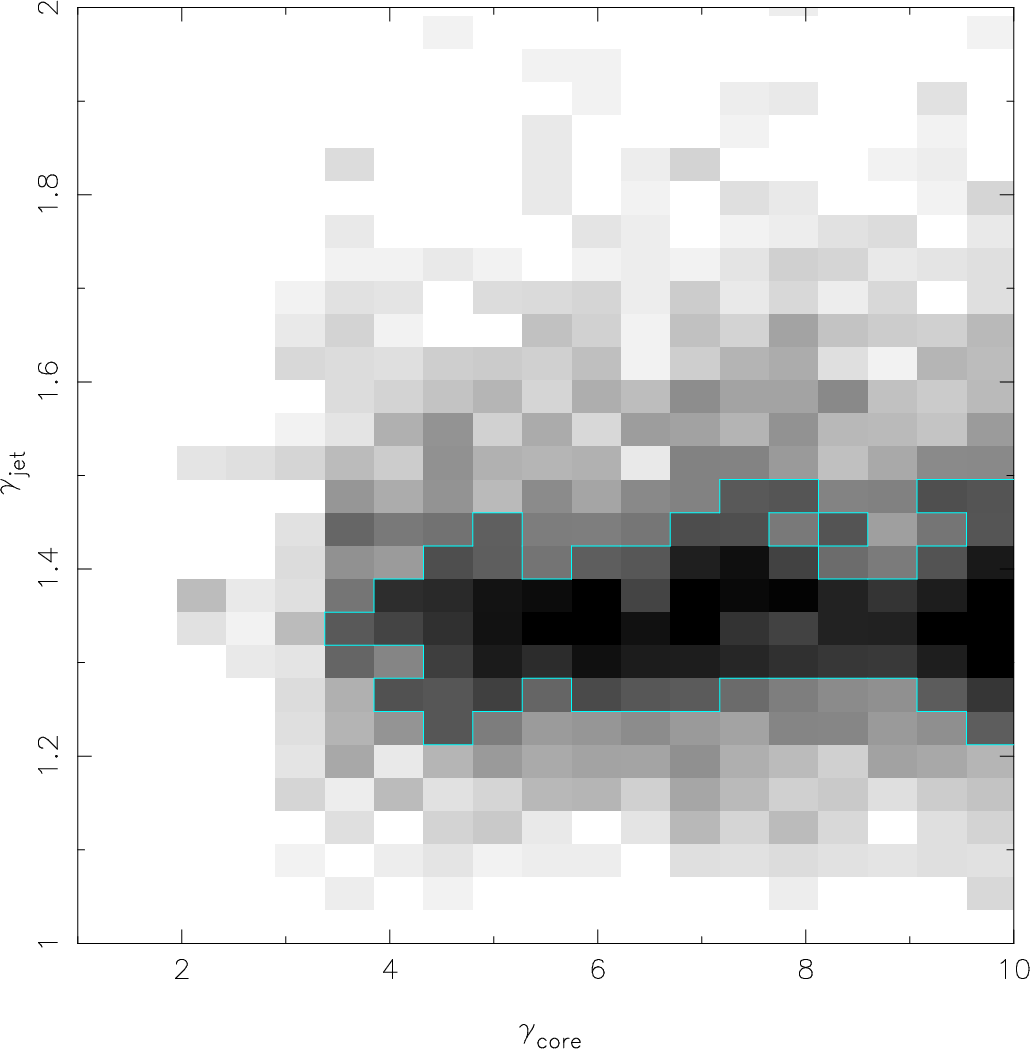}
\epsfxsize 8.5cm
\epsfbox{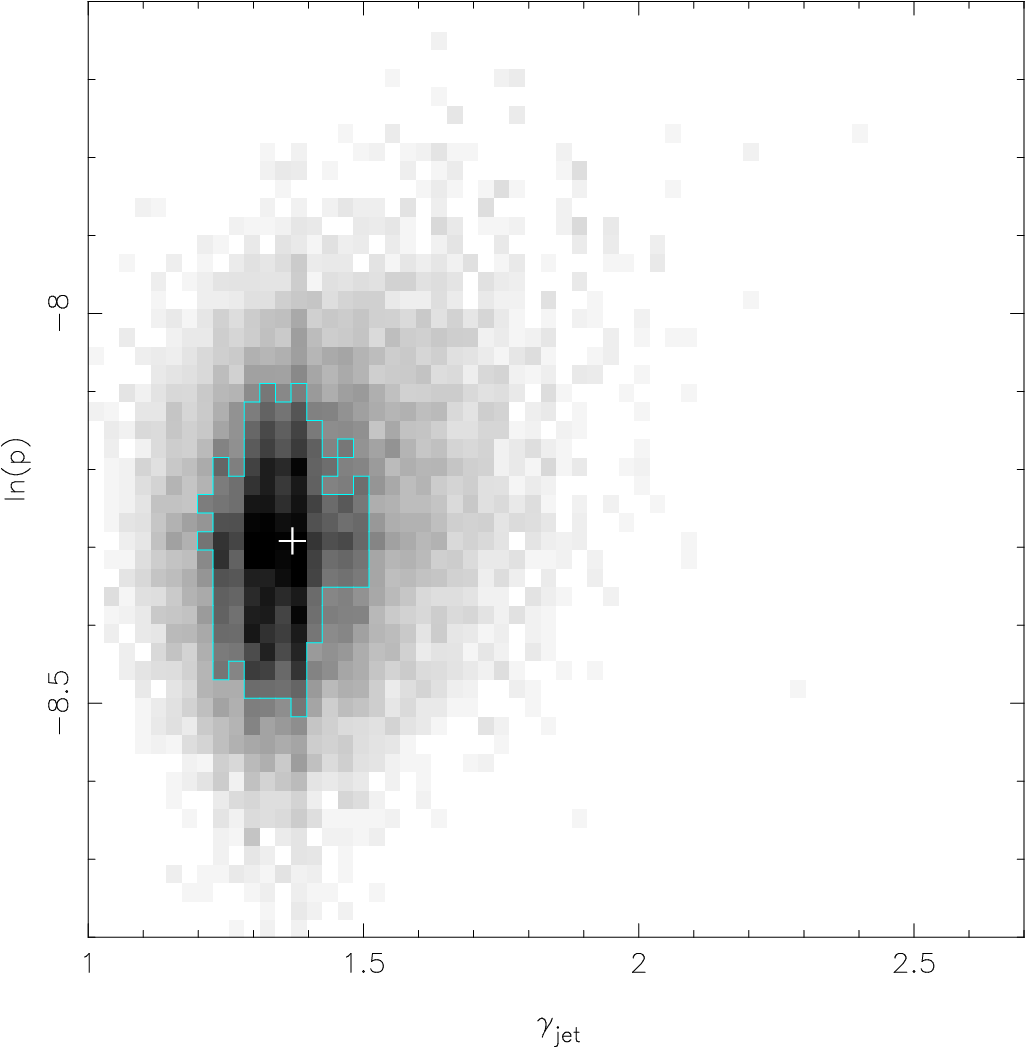}
\caption{Posterior probability distribution
  for a joint fit to the jet and core prominence distributions with
  $\alpha = 0.5$. Left:
  the beaming parameters. Right: the jet beaming and prominence
  parameters for comparison with Fig.\ \ref{fig:jet-pp}. The
  probability distributions are marginalized over all other parameters.}
\label{fig:joint-gg}
\end{figure*}

Results of these fits are tabulated in Table \ref{tab:joint-results}, 
again allowing the jet spectral index to be either 0.5 or 0.8,
and an example of the posterior probability distribution for the
beaming parameters is shown in Fig.\ \ref{fig:joint-gg}. It can be
seen that beaming is strongly required by the Bayes factor between
these models and the corresponding ones in which the Lorentz factors
are fixed to 1.0, as we might expect. The effect of including the
core-jet correlation in the modelling is to make less probable the
regions of parameter space with no beaming (compare the right-hand
panel of Fig.\ \ref{fig:joint-gg} with the top right-hand panel of
Fig.\ \ref{fig:jet-pp}). The Bayesian estimates of the beaming speeds
are therefore slightly higher than in the case where the correlation
is not taken into account ($\bar\gamma_{\rm j} = 1.35$ corresponds to
$\beta_{\rm j} = 0.67$, and $\bar\gamma_{\rm j} = 1.23$ to $\beta_{\rm j}
\approx 0.58$) although there is substantial overlap in the
credible intervals. Once again, there is little difference between the
results for all sources and the results that exclude the LERGs.

\begin{table*}
\caption{Results of joint fits to the jet and core prominence
  distributions for the whole sample for jet spectral indices $\alpha
  = 0.5$ and $\alpha = 0.8$. Conventions as for Table
  \ref{tab:jet-results}, but the Bayes factor is measured with respect
  to the model with free beaming parameters.}
\label{tab:joint-results}
\begin{tabular}{lrlrrrrrrr}
\hline
Sample&$\alpha$&Model&$\bar\gamma_{\rm c}$&$\ln(\bar p_{\rm int_c})$&$\sigma_{p_{\rm int_c}}$&$\bar\gamma_{\rm j}$&$\ln(\bar p_{\rm int_j}$)&$\sigma_{p_{\rm int_j}}$&Bayes factor\\
\hline
All sources&0.5&All free&$11.81^{+5.37}_{-4.74}$&$-5.17^{+1.10}_{-0.55}$&$1.50^{+0.12}_{-0.14}$&$1.37^{+0.06}_{-0.12}$&$-8.29^{+0.11}_{-0.14}$&$0.97^{+0.09}_{-0.11}$&--\\
&&No beaming&1.0&&&1.0&&&$-10.0$\\
&0.8&All free&$11.76^{+6.57}_{-3.55}$&$-5.18^{+1.10}_{-0.51}$&$1.49^{+0.09}_{-0.16}$&$1.23^{+0.03}_{-0.08}$&$-8.36^{+0.11}_{-0.10}$&$0.98^{+0.08}_{-0.11}$&--\\
&&No beaming&1.0&&&1.0&&&$-8.0$\\
\hline
LERGs excluded&0.5&All free&$10.49^{+3.56}_{-7.29}$&$-5.61^{+1.55}_{-0.64}$&$1.38^{+0.13}_{-0.12}$&$1.28^{+0.06}_{-0.13}$&$-8.39^{+0.13}_{-0.13}$&$1.07^{+0.06}_{-0.09}$&--\\
&&No beaming&1.0&&&1.0&&&$-4.0$\\
&0.8&All free&$9.48^{+3.27}_{-6.30}$&$-5.85^{+1.02}_{-1.56}$&$1.39^{+0.09}_{-0.12}$&$1.20^{+0.02}_{-0.05}$&$-8.41^{+0.12}_{-0.12}$&$1.05^{+0.07}_{-0.08}$&--\\
&&No beaming&1.0&&&1.0&&&$-4.2$\\
\hline
\end{tabular}
\end{table*}

%
%
\section{Discussion and Conclusions}\label{sec:conclusions}
The models for the prominences of both jets and cores are a function of three variables, ($p_{\rm int}\ \gamma,\
\theta$), so that the problem of fitting to the $\gamma$ distribution
is degenerate. However, as discussed in Section \ref{sec:modelval},
prominence distributions simulated using these these models do
represent the sample data well, and are strongly influenced by the
value of $\gamma$. The model fitting that we have performed in this paper has allowed the upper limits in the data to be treated
correctly and so the following conclusions can be drawn about the probable range of $\gamma$ in the jet and core data.

For both jets and cores, it was found that models allowing some
dispersion in one or both of the intrinsic distributions of $p_{\rm
  int}$ and $\gamma$ are favoured over the model that assumes a
delta-function distribution for both parameters. For the jets, the
value of $\bar \gamma$ found for these models varied between 1.18 and
1.49 (see Table \ref{tab:jet-results}), depending on the model
  and the choice of $\alpha$, which corresponds to $\beta \approx
0.53 - 0.74$ and is in reasonable agreement with previously reported
analysis, as discussed in Section \ref{sec:history}. Models with
$\gamma \approx 10$ did not fit well to the data and there was no
evidence for any luminosity dependence of $\bar\gamma$. For the cores,
a reasonably well constrained $\bar\gamma$ of $\approx 10$--$14$ was
obtained in models in which intrinsic dispersion in the prominence was
allowed, but the results here are dependent on our choice of prior.

One obvious criticism of the core data modeling is that it has been
assumed that the properties of the twin parsec-scale jets are the same
and that the jets are bi-polar. More complicated models could be
defined that allow the $p_{\rm int}$ and $\gamma$ values to be
different in the approaching and receding jet, and allow the
separation between the jets to be less than 180$^\circ$ -- even if the
real parsec-scale jets properties are similar or close to being
symmetrical in the two beams, it is entirely plausible there is at
least some deviation from exact symmetry. Experimenting with such
models, however, determined that the data does not justify models any
more complex than those reported in the preceding
sections and Table \ref{tab:jet-results}. That is, the evidence values
associated with the models with greater numbers of parameters are not
higher and fitting results do not differ greatly from those obtained
with the simpler models. Considerably larger datasets will be needed
to investigate more complex models.

We also carried out fits to the combined jet and core prominence
datasets, as evidence for a jet-core correlation has been found in our
sample and in others (Bridle \etal\ 1994; H99; Paper I). The results
showed that models with no beaming are strongly disfavoured with
respect to models in which beaming is included, since only the latter
can reproduce the observed core-jet correlation. The fitted $\gamma$
value for the kiloparsec-scale jets was slightly higher than for the
corresponding fits in which the correlation was not taken into account
-- $\gamma_{\rm jet} = 1.35$, which corresponds to $\beta \approx
0.67$, for $\alpha = 0.5$, or $\gamma_{\rm jet} = 1.23$, $\beta
  \approx 0.58$, for $\alpha = 0.8$. Even so, these values, which
represent our best estimates in the sense that they make use of all
the available data, are still clearly much lower than 10.

How can these results, implying mildly relativistic bulk speeds for
kpc-scale jets in FRIIs, be reconciled with the much larger values
of jet bulk Lorentz factor required by beamed inverse-Compton models
of X-ray emission from quasar jets (Section \ref{sec:history})? As we
have already pointed out, jet velocity structure is routinely invoked
in models of the parsec-scale jets in order to explain the wide range
of Lorentz factors required by different observations in those cases.
If a beamed inverse-Compton model is viable at all for
kiloparsec-scale jets, velocity structure must be present there as
well (as concluded by Hardcastle 2006). Our data are dominated by
objects at large angles to the line of sight, and so would naturally
be expected to give estimates of the bulk Lorentz factor that are
appropriate to the slow-moving component of the jet. However, the
presence of this slow-moving component in observations of jets even at
a small angle to the line of sight certainly adds to the complications
of detailed inverse-Compton modelling of FRII sources; the radio
emission from the jet is the only independent information available on
the electron density present, but the assumption that all the
radio-emitting material moves at a single bulk speed which can be
estimated from the X-rays is no longer valid in the presence of jet
velocity structure. Ideally measurements of jet prominences could be
used to constrain models with jet velocity structure, but even if our
data were good enough (and we have seen that our capability to
distinguish between models is limited only to the most obvious cases)
there is as yet no constraint on how the jet speed and emissivity
might vary as a function of radius, and so effectively no model to
test. High-resolution radio observations of jets in FRIIs with
next-generation instruments such as e-MERLIN and the EVLA are required
to give observational constraints on the surface brightness of FRII
jets at a range of angles to the line of sight as a function of jet
radius. It may then be possible to apply the techniques described
here to constrain the properties of a more complex model of jet speeds.

%
\section*{Acknowledgments}

We are very grateful to Mike Hobson for providing us with his {\sc
metro} Markov-Chain Monte Carlo code with which the initial analysis
for this project was carried out and which formed the basis of our own
implementation of the technique. We also thank an anonymous referee
for constructive comments on the first version of this paper. LMM
thanks PPARC (now STFC) for a studentship
and the astrophysics group at the Cavendish Laboratory, University of
Cambridge, for support during the early stages of this work. MJH
thanks the Royal Society for a research fellowship.

%
%

\label{lastpage}

\end{document}